\documentclass[a4paper, 11pt]{amsart}
\usepackage{graphicx}
\usepackage{amssymb}
\usepackage{amsmath}
\usepackage{rotating}
\usepackage{array,multirow}
\newcommand{\R}{\mathbb{R}}

\def\be#1\ee{\begin{equation}#1\end{equation}}
\newcommand{\fer}[1]{(\ref{#1})}

\newcommand{\bq}{\begin{equation}}
\newcommand{\eq}{\end{equation}}
\newenvironment{equations}{\equation\aligned}{\endaligned\endequation}
\def\bqa{\begin{eqnarray}}
\def\eqa{\end{eqnarray}}

\def\e{\epsilon}

\newcommand{\bd}{\begin{displaymath}}
\newcommand{\ed}{\end{displaymath}}
\newcommand{\ba}{\begin{eqnarray}}
\newcommand{\ea}{\end{eqnarray}}

\newtheorem{theorem}{Theorem}[section]

\newtheorem{remark}[theorem]{Remark}

\begin{document}

\title{Kinetic modeling of alcohol consumption}
%\markboth{G. Toscani}{Kinetic modeling of alcohol consumption}
\author{Giacomo Dimarco}

\thanks{Department of Mathematics and Informatics of the University of Ferrara
	e.mail:   giacomo.dimarco@unife.it 
}
\author{Giuseppe Toscani}

\thanks{Department of Mathematics  of the University of Pavia, and IMATI CNR, Italy.  e.mail:  giuseppe.toscani@unipv.it. 
}

\date{\today}

\maketitle

\begin{abstract}
In most countries, alcohol consumption distributions have been shown to possess universal features. Their unimodal right-skewed shape is usually modeled in terms of the Lognormal distribution, which is easy to fit, test, and modify. However,  empirical distributions often deviate considerably from the Lognormal model, and both Gamma and Weibull distributions appear to better describe the survey data. In this paper we explain the appearance of these distributions  by means of classical methods of kinetic theory of multi-agent systems. 
The microscopic variation of alcohol consumption of agents around a universal \emph{social} accepted value of consumption, is  built up introducing as main criterion for consumption  a suitable value function in the spirit of the prospect theory of Kahneman and Twersky. The mathematical properties of the value function then determine the unique macroscopic equilibrium  which results to be a generalized Gamma distribution. The modeling of the microscopic kinetic interaction allows to clarify the meaning of the various parameters characterizing the 
generalized Gamma equilibrium.
%\subclass{35Q84\and 82B21\and 91D10\and 94A17}
\end{abstract}
\section{Introduction}

Alcohol consumption is a component cause for a number of diseases, which makes studies on it very relevant from both medical and psychological point of views. Information about this social phenomenon requires to know at best its statistical distribution, to analyze the effect of the resulting alcohol consumption distribution on the estimation of the alcohol Population-Attributable Fractions, and to characterize the chosen alcohol consumption distribution by exploring if there is a global relationship within the distributions \cite{Keh,Reh}.

Despite the increasing number of papers devoted to a precise fitting of the statistical distribution of the phenomenon of alcohol consumption and to the social motivations about it, the precise connections leading from the human behavior of alcohol consumers to the observed  statistical distributions have been left largely unexplored, and a satisfactory mathematical modeling is still missing. 

We started to care of this problem after the reading of two interesting research papers recently published into the journal \emph{Population Health Metrics} \cite{Keh,Reh}, whose conclusions were subsequently considered as a starting point for further statistical studies (cf. \cite{Par} and the references therein). In these papers the analysis of the fitting of real data about alcohol consumption in a huge number of countries, led the authors to the conclusion that, among various probability distributions often used in this context, Gamma and Weibull distributions appeared to furnish a better fitting with respect to the Lognormal distribution, usually used in this context, starting from the work of Ledermann \cite{Led}, who proposed it as a reasonable model for the consumption problem.

Concerning the mathematical modeling of alcohol consumption by resorting to the methods of statistical physics, the results of the analysis presented in \cite{Keh,Reh} introduce into the problem of characterizing its distribution a number of challenging questions. Given a system of agents characterized in time by their level of alcohol consumption, and given a certain observed statistical distribution of their consumption,  can we describe at the level of the single agents the stylized facts which lead at a macroscopic level to this distribution? And moreover, can we characterize at the level of single agents the differences which produce a variety of steady  probability distributions for the same problem? 

As far as the Lognormal distribution is concerned, the kinetic modeling of phenomena concerned with interacting multi-agent systems and leading to this distribution has been recently proposed in \cite{GT18}, starting from the results of \cite{GT17}. 

Among probability distributions on $\R_+$, Lognormal distribution has a leading role \cite{Aic,Cro}. The 
list of phenomena which fit Lognormal distribution in natural sciences is quite long, as documented the exhaustive review paper by Limpert, Stahel and Abbt \cite{Lim}. Moreover, the success of Lognormal distribution in the study of random variations that occur in the data from these phenomena that show more or less skewed probability distributions, is favored by its features, which render it easy to fit, test and modify.   

In \cite{GT18}, it was pointed out that a relevant number of phenomena involving measurable quantities of a population and fitting Lognormal distribution comes from social sciences and economics, areas where it can be reasonably assumed that the appearance of this distribution is a consequence of a certain human behavior.  

Among others, a good fitting has been observed while looking at the distribution of body weight \cite{BC}, at women's age at first marriage \cite{Pre}, at drivers behavior \cite{JJ}, or, from the economic world, when looking at consumption in a western society \cite{BBL}, at the size of cities \cite{BRS}, and at call-center service times \cite{Brown}. 

These phenomena were modelled in \cite{GT18} by resorting to the methodology of classical kinetic theory, and  the Lognormal distribution (the macroscopic equilibrium of the multi-agent system)  appears as a direct consequence of the details of the microscopic interactions of agents. This idea is fully coherent with the classical kinetic theory of rarefied gases, where the formation of a Maxwellian equilibrium in the spatially uniform Boltzmann equation is closely related to the conservation laws of the elastic microscopic binary collisions between molecules \cite{Cer,GT-ec}. 

The analysis of \cite{GT17,GT18} suggests a way to modify the microscopic interaction of the variation of the alcohol consumption of agents, that leads to the class of distributions fitted in \cite{Keh,Reh}. As in \cite{GT17,GT18}, the elementary single change will be expressed in terms of a suitable value function, that meets the same properties as the one leading to the Lognormal distribution, built up in the spirit of the prospect theory of Kahneman and Twersky \cite{KT,KT1}. As usual in the kinetic setting \cite{FPTT}, the microscopic interaction generates a linear kinetic equation for the density of agents, that will be subsequently studied in the asymptotic regime of \emph{grazing} interactions \cite{Vi}. 
The  \emph{grazing} regime describes the situation in which a single interaction produces only a very small change of the consumption variable. In this regime, the variation of density of the alcohol consumption of the agent's system is driven by a partial differential equation of Fokker--Planck type \cite{FPTT}.

The precise mathematical model is the following. Let us denote by $g = g(w,t)$ the density of agents which have a (positive) level of consumption equal to $w$ at time $t\ge 0$. Then, the density $g$ is the solution of a linear Fokker--Planck equation with variable coefficients of diffusion and drift, given by
\begin{equation}\label{FPori}
 \frac{\partial g(w,t)}{\partial t} = \frac \lambda 2 \frac{\partial^2 }{\partial w^2}
 \left(w^2 g(w,t)\right )+ \frac \mu{2}
 \frac{\partial}{\partial w}\left[ \frac 1\delta\left( \left( \frac w{\bar w_L} \right)^\delta -1 \right)w g(w,t)\right].
 \end{equation}
In \fer{FPori}  $\bar w_L$ represents the critical (maximal) level of individual alcohol consumption in the chosen time unity, while $1-\delta$, with $0\le \delta \le 1$ is linked to the level of psychophysical addiction induced by alcohol . Moreover,  $\lambda$ and $\mu$   are positive constants  closely related to the randomness and intensity of the phenomenon under study, satisfying further bounds imposed by the physical conditions of the microscopic interaction. In particular, if $\gamma = \mu/\lambda$, $ \gamma >\delta$, and the equilibrium density of the Fokker--Planck equation \fer{FPori} is given by the (non normalized) generalized Gamma density  \cite{Lie,Sta}
 \be\label{equili}
g_\infty(w) = g_\infty(\bar w_L)\left( \frac w{\bar w_L} \right)^{\gamma/\delta -2}  \exp\left\{ - \frac \gamma{\delta^2}\left( \left( \frac w{\bar w_L} \right)^\delta -1 \right)\right\},
 \ee 
Moreover, for any given initial distribution $g_0(w)$ of agents, convergence to  equilibrium can be shown to hold exponentially fast in time with explicit rate \cite{OV}. 
Note that  the steady distribution \fer{equili} includes as particular cases the Gamma distribution, and in this case $\delta =1$, as well as the Weibull distribution, and in this other case $\gamma = \delta(1+\delta)$. Moreover, a (non normalized) Lognormal distribution
\be\label{LN}
g_\infty(w) = g_\infty(\bar w_L)\left( \frac w{\bar w_L} \right)^{- 2}  \exp\left\{ - \frac \gamma{2}\left( \left( \log\frac w{\bar w_L} \right)^2 \right)\right\},
\ee
is included into \fer{equili} and it is obtained from \fer{equili} by taking the limit $\delta \to 0$. Hence, one can conjecture that the best fitting of the alcohol consumption distribution is achieved by suitably choosing the three parameters characterizing the generalized Gamma distribution, namely $\gamma, \delta$ and $\bar w_L$.   

It is remarkable that the fitting analysis of \cite{Keh,Reh}, was limited to the choice of particular cases of the generalized Gamma distribution \fer{equili},  obtained by   fixing precise values of the parameter $\delta$.  Hence, in \cite{Keh,Reh} the fitted distributions are characterized by the presence of only two  parameters.

In the kinetic description leading to the steady state \fer{equili}, one of these parameters is given by $\bar w_L$, representing the critical (maximal) level of individual alcohol consumption. Following Ledermann \cite{Led}, this parameter can be fixed a priori, thus reducing  the two-parameter  distributions considered in \cite{Keh,Reh} to  one-parameter distributions. This generates a new way of fitting that follows along the lines indicated by Ledermann. In the case of the Lognormal distribution the description of the variation of the Lognormal distribution with respect to the unique unknown parameter is given by the so-called \emph{Ledermann curve} \cite{Led}, which describes the percentage of consumers drinking above a fixed level. Analogous results can be derived in the other cases.

The mathematical methodology of this paper takes essential advantage from previous attempts to model social and economic phenomena in a multi-agent system.  Starting from economics, where the genesis of the formation of Pareto curves in wealth distribution of western countries was studied in various aspects \cite{ChaCha00,CCM,ChChSt05,CoPaTo05,DY00,GSV,SGD}, this field of research moved to social sciences, where the investigation of opinion formation played a leading rule \cite{BN2,BN3,BN1,BeDe,Bou,Bou1,Bou2,CDT,DMPW,GGS,GM,Gal,GZ,SW,To1}. 

Recently, some of these investigations introduced into the microscopic description of the interactions specific behavioral aspects of agents. To our knowledge, this idea has been first proposed by Pareschi and Maldarella in \cite{MD}. There the authors, to investigate the price formation of a good in a multi-agent market, introduced a kinetic model for a multi-agent system consisting of  two different trader populations, playing different rules of trading, and possibly changing their point of view. The kinetic description was inspired by the microscopic Lux--Marchesi model \cite{LMa,LMb} (cf. also \cite{LLS,LLSb}).  In \cite{MD}  the trading rules of agents were assumed to depend on the opinion of traders through a kinetic model of opinion formation recently introduced in \cite{To1}. Also, psychological and behavioral components of the agents, like the way they interact with each other and perceive risks, were taken into account. This has been done by resorting, in agreement with the prospect theory by Kahneman and Twersky \cite{KT,KT1}, to interactions containing a suitable \emph{value function}. 

The analysis of \cite{MD} enlightened in a kinetic description the role of  human behaviors \cite{BHT,BKS,BCKS}, pioneered in \cite{Zipf}. These aspects have been recently considered in \cite{GT17}, where the introduction of a suitable value function justified at a microscopic level the mechanism of formation of the service time distribution in a call center. 

The leading idea in \cite{GT17}  is based on a general principle which can be easily verified in a number of social activities of agents in which one identifies the possibility of a certain addiction. 

The paper is organized as follows. In Section \ref{model}, we shall introduce a linear kinetic model for a multi-agent system, in which agents are characterized in terms of  their level of alcohol consumption that can be measured in terms of a fixed unit of measure, and subject to microscopic interactions which describe the microscopic rate of change of the value of their consumption, according to the previous general principle. 
The relevant mechanism of the microscopic interaction is based on a suitable value function, in the spirit of the analysis of Kahneman and Twersky \cite{KT,KT1}, which reproduces at best the human behavior in this situation.

In Section \ref{quasi} we will show that in a suitable asymptotic procedure (hereafter called \emph{grazing}  limit) the solution to the kinetic model tends towards the solution of the Fokker-Planck type equation \fer{FPori}. Similar asymptotic analysis was performed in \cite{CPP,DMTb} for  a kinetic model for the distribution of wealth in a simple market economy subject to microscopic binary trades in presence of risk, showing formation of steady states with Pareto tails, in \cite{TBD} on kinetic equations for price formation, and in \cite{To1} in the context of opinion formation in presence of self-thinking. A general view about this asymptotic passage from  kinetic equations based on general interactions  towards Fokker--Planck type equations can be found in \cite{FPTT}. Other relationships of this asymptotic procedure with the classical problem of the \emph{grazing collision limit} of the Boltzmann equation in kinetic theory of rarefied gases have been recently enlightened in \cite{GT-ec}.

Once the  Fokker--Planck equation \fer{FPori} has been derived, in Section \fer{numerics} we will investigate at a numerical level the relationships  between the solutions of the linear kinetic model of Boltzmann type and its \emph{grazing} asymptotics in the form of the Fokker--Planck equation, at different values of the vanishing $\e$-parameter leading to the asymptotic itself. The numerical computations will confirm the perfect agreement between the stationary solutions of the two models, even for values of the $\e$-parameter of the order $10^{-1}$. This justifies, at least numerically, the role of the chosen value functions in the asymptotic description of the alcohol consumption phenomenon. 

%%%%%%%%%%%%%%%%%%%%%%%%%%%%%%%%%%%%%%%%%%%%%%%%%%%%%%%%%%%%%%%%%

%%%%%%%%%%%%%%%%%%%%%%%%%%%%%%%%%%%%%%%%%%%%%%%%%%%%%%%%%%%%%%%%%
\section{A linear Boltzmann equation}\label{model}
\subsection{Modeling assumptions}
\label{sec:1.1}
Following the well-consolidated approach furnished by the kinetic theory \cite{FPTT,NPT,PT13}, the statistical description of alcohol consumption in a society will be described by resorting to a linear Boltzmann-type equation in which the unknown is the probability distribution of alcohol consumption of the agent system. Since we want to fit the results obtained by this approach with the various dataset coming from real situations, the kinetic model is built up to respect some basic hypothesis we enumerate below. 

First of all, given a population of agents,  the population needs to be considered homogeneous with respect to alcohol consumption. In this case, to have a homogenous population, it is essential to restrict it with respect to some characteristics, like age, sex and social class \cite{Led,Skog}. 
Once the homogeneity assumption is satisfied, agents in the system are considered indistinguishable \cite{PT13}. This means that an agent's state at any instant of time $t\ge 0$ is completely characterized by the level $w \ge0$ of its alcohol consumption.  This level can be easily and uniformly measured with respect to some unit. For example, $w$ indicates the mean value of the daily consumption of alcohol in grams of agents.
The unknown is the density (or distribution function) $f = f(w, t)$, where $w\in \R_+$ and the time $t\ge 0$, and the target is to study its time evolution towards a certain equilibrium.

The precise meaning of the density $f$ is the following. Given the system of agents to study, and given an interval or a more complex sub-domain $D  \subseteq \R_+$, the integral
\[
\int_D f(w, t)\, dw
\]
represents the number of individuals with  are characterized by a level $w \in D$ of consumption at time $t > 0$. It is assumed that the density function is normalized to one, that is
\[
\int_{\R_+} f(w, t)\, dw = 1.
\]
The change in time of the density is due to the fact that agents continuously upgrade the level $w$ of their alcohol consumption in time by some action.  To maintain the connection with classical kinetic theory of rarefied gases, we will always refer to a single upgrade of the level measure as a microscopic \emph{interaction}. 

\subsection{The value function in alcohol consumption}
\label{sec:1.2}
Similarly to the problems treated in \cite{GT17,GT18}, alcohol consumption deeply depends on both addiction and social attitudes, that can be easily translated by saying that, at least over a certain level, agents likely tend to increase the value $w$ by interactions, while manifest a certain resistance to decrease it. 

For the situation under study we can identify (and fix) for the system of agents two \emph{universal} values. The first value, denoted by $\bar w$, is the \emph{social} barrier, and identifies the mean amount of alcohol consumption relative to the homogenous class, which is normally accepted by the social rules of the underlying society. The second value,  denoted by  $\bar w_L$, with $\bar w_L > \bar w$, is the threshold value that it would be better not to exceed to enter into addiction.  Consequently, the human tendency to increase the value $w$ by interactions has to be coupled with the existence of this limit value $\bar w_L$ which serious health  reasons suggest not to cross. Similar  behaviors have been recently modeled in \cite{GT17,GT18} by resorting to the prospect theory introduced by Kahneman and Twersky  in their pioneering paper \cite{KT}, devoted to describe decision under risk. 

Following \cite{GT17,GT18}, we will describe the microscopic variation of alcohol consumption of agents in the form
\be\label{coll}
w_* = w  - \Phi(w/\bar w_L) w + \eta w.
\ee
According to \fer{coll}, in a single interaction the value $w$ of alcohol consumption can vary by two different motivations. The first one determines a variation of a  quantity proportional to $w$ in which the coefficient $\Phi(\cdot)$, which can assume assume both positive and negative values, is a function characterizing the predictable behavior of agents. The second one, still proportional to $w$, takes into account a certain amount of human unpredictability. The random variable $\eta$ describes the unpredictable variations,  which in the mean are assumed negligible, and in any case are not so significant to produce a sensible change of the value $w$.  Hence, the consumption of alcohol can be both increasing and decreasing, and the mean intensity of this variation is fully determined by the function $\Phi$. 

The function $\Phi$ plays the role of the \emph{value function} in the prospect theory of Kahneman and Twersky \cite{KT}, and contains the mathematical translation of the expected human behavior in the phenomenon under consideration. 

In their pioneering theory \cite{KT,KT1}, Kahneman and Twersky, by resorting to various situation concerned with decision theory, identified the main properties characterizing a value function. This function has to be positive and concave above the reference value $1$ ($w > \bar w_L$), while negative and convex below ($w < \bar w_L$). Hence, in terms of the variable $s = w/\bar w_L $, provided that the value function is enough regular in $s$ \be\label{ccc}
\Phi''(s) \le 0 \quad 0\le s< 1, \qquad \Phi''(s) \ge 0 \quad s> 1.
\ee
Thus, as far as financial transactions are considered, the value function is generally concave for gains and commonly convex for losses. Also, by noticing that the aversiveness of symmetric fair bets generally increases with the size of the stack, they assumed that the value function satisfies the further properties
\be\label{ccd}
- \Phi\left(1-s \right) > \Phi\left(1+s \right),
\ee
and 
\be\label{cce}
\Phi'\left(1+s \right)< \Phi'\left(1-s \right). 
\ee
Hence, properties of value functions are well defined for deviations from the reference point ($s=1$ in our case), and the value function for losses is steeper than the value function for gains. 

In \cite{GT17} we adapted the shape of the value function given in \cite{KT} to the problem of the characterization of the distribution of the length of a service time in a call center. A careful reading of \cite{GT17} allows to recognize that  the properties of the value function do not have a universal validity, since its precise form  depends on the main aspects of the human behavior relative to the phenomenon under study.  At difference with the choice of Kahneman and Twersky, in \cite{GT17} a suitable  value function  was identified in the increasing concave function
\be\label{vf}
\Phi_0^\e(s) = \mu \frac{s^\e -1}{s^\e +1} , \quad  s \ge 0.
\ee
In \fer{vf} $0<\mu < 1$ and $0< \e < 1$ are suitable constants characterizing the agents behavior. In particular, the value $\mu$ denotes the maximal amount of variation of service time that agents will be able to obtain in a single interaction. Note indeed that the value function $\Phi_0(s)$ is such that 
\be\label{bounds}
-\mu \le \Phi_0^\e(s) \le \mu.
\ee
The function in \fer{vf} satisfies properties \fer{ccd} and \fer{cce}, but in consequence of its uniform concavity does not satisfy assumption \fer{ccc}. Note that in the service time problem studied in \cite{GT17}, property \fer{ccd} characterizes in a proper way the behavior of agents which have to do a service by respecting a certain time. Indeed,  given  two agents starting at the same distance from the limit value $\bar w_L$ from below and above, it will be easier from the agent starting below to move closer to the optimal time, than for the agent starting above.

Going back to the form of the microscopic interaction \fer{coll}, the minus sign in front of the value function $\Phi$ reflects the tendency of agents to increase their value $w$ of alcohol consumption  when $w < \bar w_L$,  while they would like to decrease consumption if $w >\bar w_L$. 

As shown in \cite{GT17,GT18}, the choice of the value function \fer{vf}  leads, as explained in details in the forthcoming Section \ref{quasi}, to a stationary distribution in the form of a Lognormal distribution. Hence, this choice would be in full agreement with the data fitting of alcohol consumption proposed by Ledermann in 1956 \cite{Led}, choice which is still used in present times (cf. the recent paper \cite{Mie} and the references therein).

On the other hand, as briefly discussed in the introduction, the recent fitting analysis of alcohol consumption in \cite{Keh,Reh} seems to corroborate the hypothesis that, with respect to the fitting provided through a Lognormal distribution, a better fitting can be obtained by resorting to Gamma and Weibull distributions. These results pose a question about the correctness of the shape of the value function \fer{vf}, which could fail to represent at best the human behavior relative to alcohol consumption. 

One of the points in which the shape of the value function $\Phi_0^\e(s)$, defined in \fer{vf}, could be not so well-adapted to the description of alcohol consumers, is its behavior for small values. Indeed, the first derivative of $\Phi_0(s)$ in $s=0$ is blowing up. Hence, agents with a very low level of alcohol consumption are pushed very easily towards a higher consumption, action that does not agree with the common behavior of low drinkers. Indeed, agents which are characterized by a level $w \ll w_L$, usually change their alcohol habits very slowly. Further,  the lower bound in \fer{bounds} does not depend on the value of the critical consumption value $\bar w_L$.  Hence, in terms of $\Phi_0(s)$ the maximal allowed increase in alcohol consumption in a single interaction does not depend on the distance of the point $s$ from the reference point, which does not appear to be supported from previous studies \cite{Skog}.

For these reasons, for the description of alcohol consumption, the value function 
\be\label{v1}
\Phi_1^\e(s) = \mu \frac{e^{\e(s -1)}-1}{e^{\e(s -1)}+1 } , \quad  s \ge 0,
\ee
seems more adapted. Similarly to \fer{vf}, in \fer{v1}  $0<\mu < 1$ and $0< \e < 1$ are suitable constants characterizing the intensity of interactions of agents. At difference with the value function \fer{vf}, the function \fer{v1} satisfies all properties originally proposed by Kahneman and Twersky in \cite{KT}, including the property to be positive and concave above the reference value ($s >1 $), while negative and convex below ($s <1$). Moreover, at difference with \fer{vf}, while the upper bound does not depend on $\e$, the lower bound depends on it, and 
\be\label{lbb}
\Phi_1^\e(s) \ge - \mu \,\frac{1 -e^{-\e}}{1 + e^{-\e}}.
\ee
This property implies that, for small values of the parameter $\e$, the maximal amount of increase of the level of alcohol consumption is of the order of $\e$.  

Together with \fer{v1} and \fer{vf} we consider the class of value functions 
\be\label{vd}
\Phi_\delta^\e(s) = \mu \frac{e^{\e(s^\delta -1)/\delta}-1}{e^{\e(s^\delta -1)/\delta}+1 } , \quad  s \ge 0,
\ee
characterized by the presence of a parameter $0 < \delta <1$, that interpolate between $\Phi_0^\e(s)$ and $\Phi_1^\e(s)$. The meaning of the $\delta$ parameter can be obtained by noticing that, for fixed values of $s$ and $\e$, the value function \fer{vd}, considered as function of $\delta$, is increasing. Hence, for values of $s<1$, which are such that $ \Phi_\delta^\e(s) <0$, the function $|\Phi_\delta^\e(s)|$ decreases with $\delta$. For example,  at the point $s=0$, given a fixed value $\bar\e$ of $\e$, we obtains
\be\label{vd-0}
\Phi_\delta^{\bar\e}(0) = -D(\delta) = - \mu \frac{1 - e^{-\bar\e/\delta}}{1 + e^{-\bar\e/\delta} } , 
\ee
where the function $D(\delta)$ is decreasing from the value $\mu$, assumed for $\delta \to 0$. Looking at the action of the value function in the elementary interaction \fer{coll}, it appears evident that the interval in which the value function plays a major role  is the interval $0\le s \le 1$, where we are taking into account agents that are increasing their consume of alcohol, starting from $w=0$. Hence, the value of the parameter $\delta$ is essentially linked to the greater or lesser degree of dependence exerted by the consumption of alcohol relative to agents characterized by a medium-low consumption. The previous discussion clarifies that small values of the parameter $\delta$ correspond to high values of increasing consumption and viceversa. In other words, values of $\delta$ close to zero represent the situation in which it is easier to increase the value of alcohol consumption starting from a low level.

\begin{figure}\centering
	{\includegraphics[width=6.5cm]{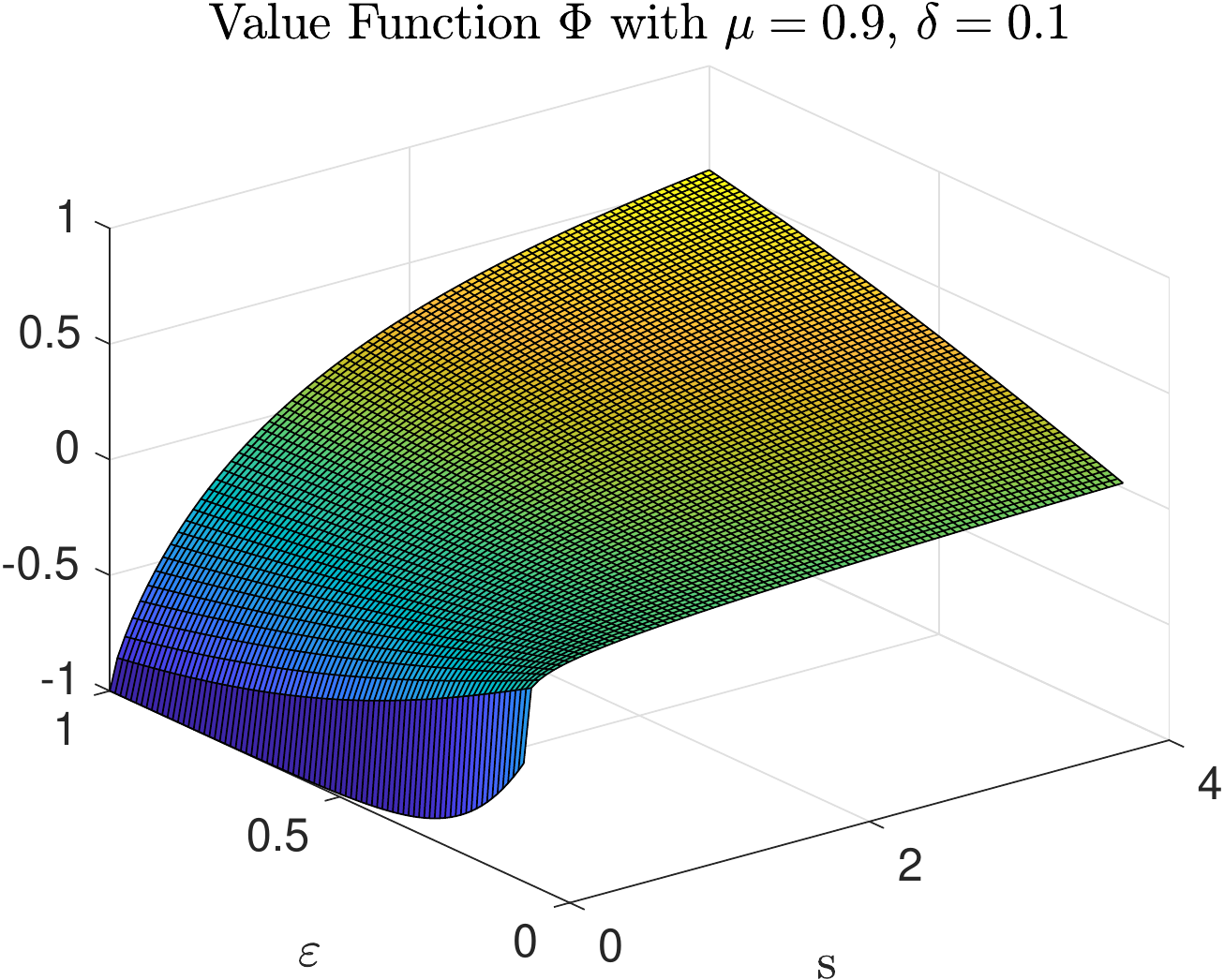}}
	\hspace{+0.35cm}
	{\includegraphics[width=6.5cm]{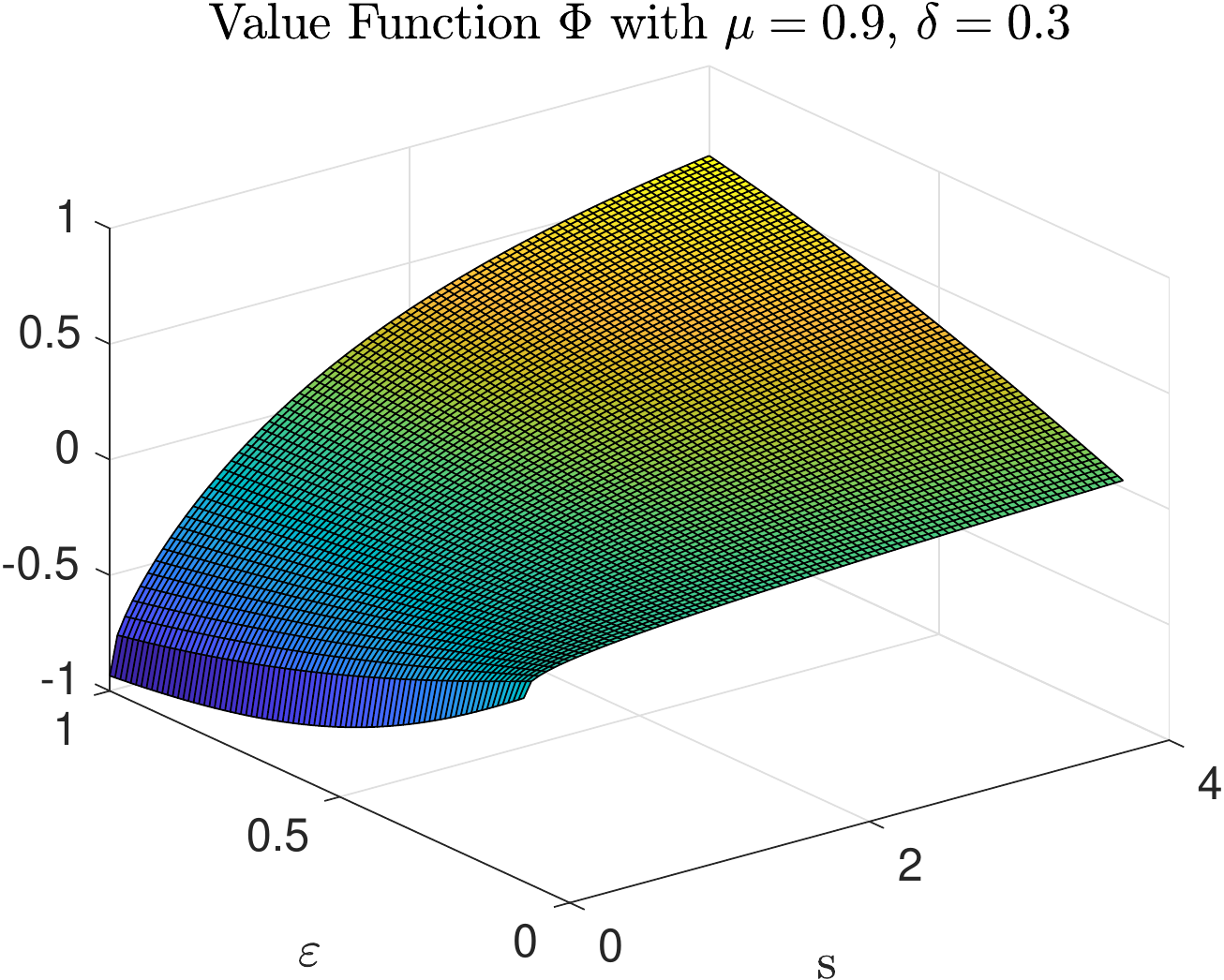}}\\
	\vspace{+0.45cm}
	{\includegraphics[width=6.5cm]{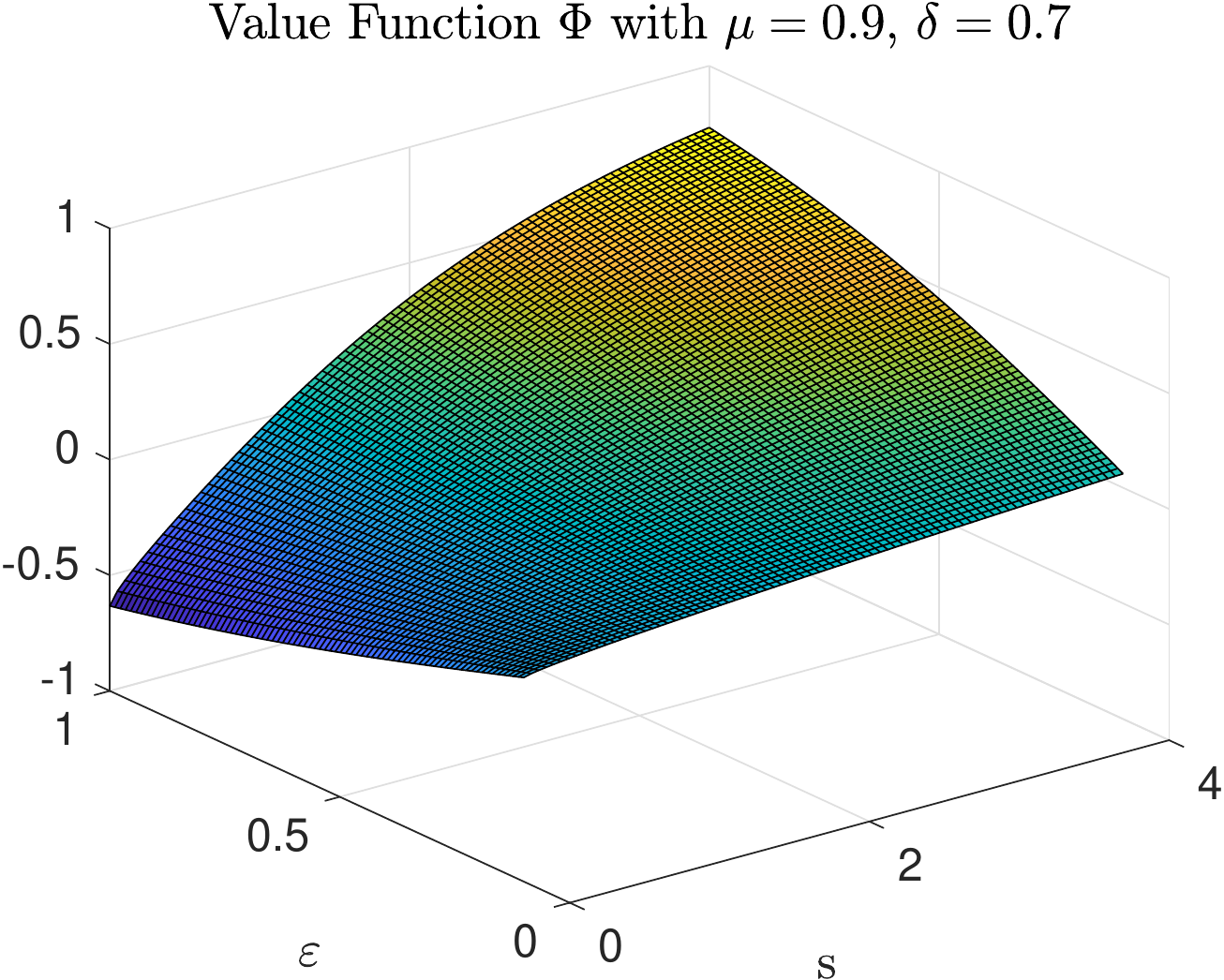}}
	\vspace{+0.35cm}
	{\includegraphics[width=6.5cm]{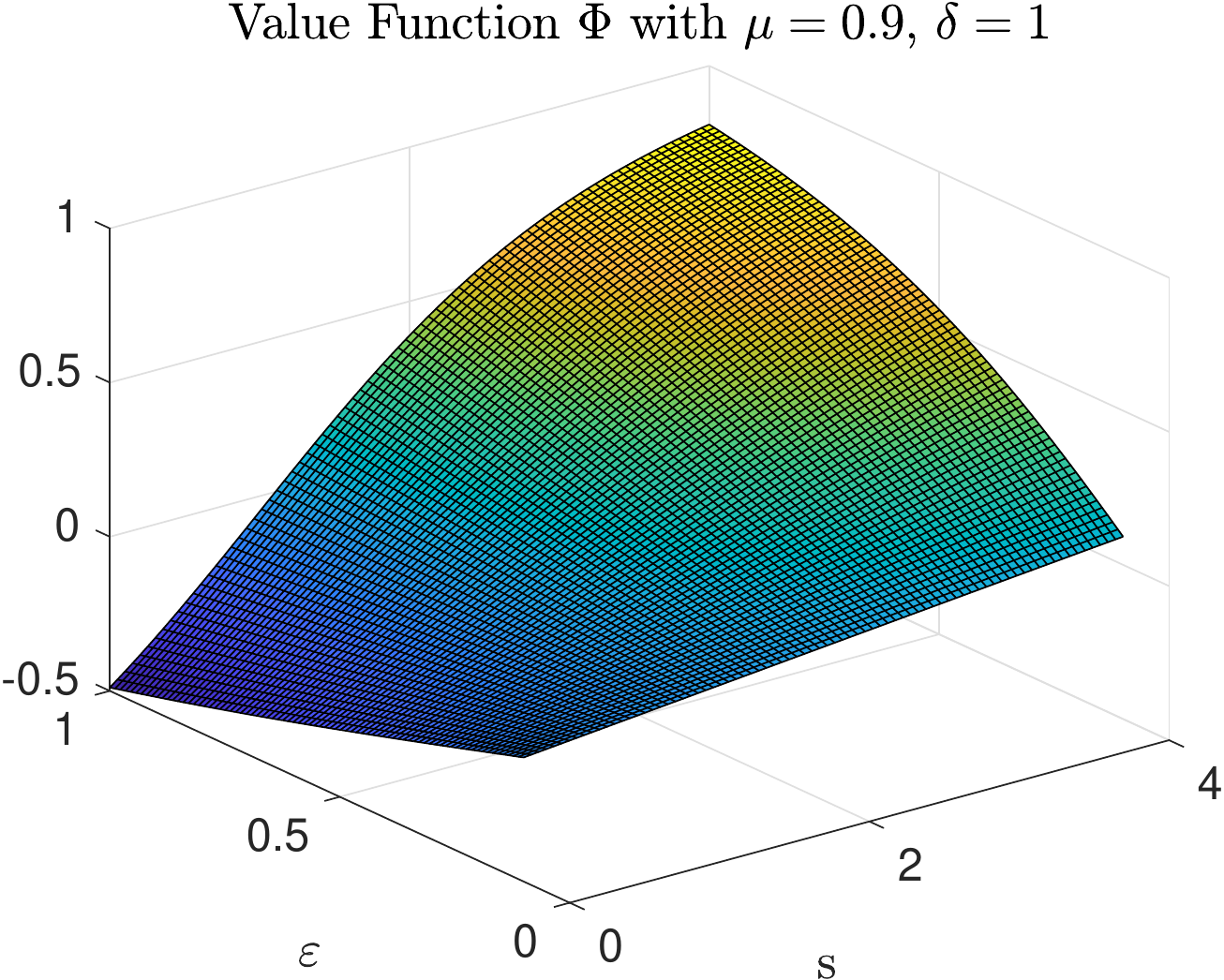}}
	\caption{Representations of the value functions $\Phi_\delta^\e(s)$ for different choices of the shape factor $\delta$ and for fixed $\mu=0.9$. }\label{fig:value_functions}
\end{figure}

In Figure \ref{fig:value_functions} we report the shapes of $ \Phi_\delta^\e(s) $ for different choices of the parameter $\delta$. 
As proven in the appendix, these functions satisfy properties \fer{ccd} and \fer{cce}, which render them suitable as value functions.

\subsection{The kinetic model}
Given the elementary \emph{interaction}  \fer{coll}, and fixed a value function $\Phi$, the study of the time-evolution of the distribution of the level of alcohol consumption follows by
resorting to kinetic collision-like models \cite{Cer,PT13}. The variation of the  density $f(w,t)$  obeys to a linear
Boltzmann-like equation, fruitfully written
in weak form. The weak form corresponds to say that the solution $f(w,t)$
satisfies, for all smooth functions $\varphi(w)$ (the observable quantities) the integro-differential equation
\begin{equation}
\label{kin-w}
\frac{d}{dt}\int_{\R_+}\varphi(w)\,f(w,t)\,dx  = \frac 1\tau
\Big \langle \int_{\R_+} \bigl( \varphi(w_*)-\varphi(w) \bigr) f(w,t)
\,dw \Big \rangle.
\end{equation}
In \fer{kin-w} the expectation $\langle \cdot \rangle$ takes into account the presence of the random parameter $\eta$ in \fer{coll}. The positive constant $\tau$ measures the interaction frequency.

The right-hand side of equation \fer{kin-w} represents the variation in density between agents that modify their value from $w$ to $w_* $ (loss term with negative sign) and   agents  that  change their value from  $w_*$ to  $w$  (gain term with positive sign).

In reason of the nonlinearity (in the variable $w$) of the interaction \fer{coll}, one easily verifies that  equation \fer{kin-w} possesses only one conserved quantity, which is obtained by setting $\varphi = 1$. This conservation law implies that the solution to \fer{kin-w} departing from a probability density function remains a probability density for all subsequent times $t >0$. Note that, provided the initial density has a certain number of moments bounded, their evolution  is difficult to follow analytically. However, as shown in \cite{GT17}, one can obtain time-dependent upper bounds for the mean value, and for higher moments. These bounds depend of the analogous bounds for the value function, as given by \fer{bounds}. These bounds do not change by changing the value of the parameter $\delta$.

  %%%%%%%%%%%%%%%%%%%%%%%%%%%%%%%%%%%%%%%%%%%%%%%%%%%%%%%%%
\section{Grazing limit and the Fokker-Planck equation}\label{quasi}

%%%%%%%%%%%%%%%%%%%%%%%%%%%%%%%%%%%%%%%%%%%%%%%%%%%%%%%%%%%%%%%
\subsection{The grazing  limit}

The linear kinetic equation \fer{kin-w} describes the evolution of the density consequent to interactions of type \fer{coll}, and it is valid for any choice of the parameters $\delta, \mu$ and $\lambda$. In real situations, however, it happens that a single interaction  determines only an extremely small change of the value $w$. This situation is well-known in kinetic theory of rarefied gases, where interactions of this type are called \emph{grazing} collisions \cite{PT13,Vi}.  In the value functions \fer{vd} the smallness assumption requires to fix $\e \ll1$. At the same time \cite{FPTT}, the balance of this smallness with the random part is achieved by setting 
\be\label{scal}
\eta \to \sqrt\e \eta.
\ee
In this way the scaling assumptions allow to retain the effect of all parameters in \fer{coll} in the  limit procedure. An exhaustive discussion on these scaling assumptions can be found in \cite{FPTT} (cf. also \cite{GT18} for analogous computations in the case of the Lognormal distribution). For these reasons, we address the interested reader  to these review papers for details. In what follows we only outline the differences in the asymptotic procedure induced by the new value functions \fer{vd}. Using \fer{scal}, we write the evolution of the mean value 
\[
m(t) = \int_{\R_+} w \, f(w,t) \, dw
\]
in the form
\[
\frac{d }{dt}\, m(t) = \e \bar w_L \, \, \frac 1\tau\int_{\R_+} \frac 1\e \Phi_\delta^\e\left(w/\bar w_L\right) \, \frac w{\bar w_L}\, \, f(w,t)\, dw.
\]
It follows from Lagrange theorem  that, if $s \ge 1$
\[
\frac 1\e \Phi_\delta^\e\left(s\right) \le \mu( s-1),
\]
and, for $s \le 1$
\[
\frac 1\e \Phi_\delta^\e\left(s\right)s \ge - \mu.
\] 
These bounds are identical to the ones obtained in \cite{GT17} for $\Phi_0^\e$. Hence, provided that the second moment of the initial density is bounded, the scaling \fer{scal} is such that, for any given fixed time $t >0$, the consequent variation of the mean value $m(t)$ is small with $\e$ small.  To observe an evolution of the average value independent of $\e$, it is enough to resort to a scaling of the frequency $\tau$. If we set 
\be\label{scal2}
\tau \to \e \tau ,
\ee
and $f_\e(w, t) $ will denote the density corresponding to the scaled interaction and frequency, then  the evolution of the average value for $f_\e(w, t)$ satisfies
\[
\frac{d}{dt}\int_{\R_+}w \,f_\e(w,t)\,dw  =  \bar w_L\, \, \frac{1}\tau\int_{\R_+}\,\frac 1\e \Phi_\delta^\e\left(w/\bar w_L\right) \, \frac w{\bar w_L}\,\, f_\e(w,t)\, dw, 
\]
namely a computable evolution law for the average value of $f$, which remains bounded even in the limit $\e \to 0$, 
since pointwise
\be\label{AA}
A_{\delta,\e}\left(w\right) = \frac 1\e \Phi_\delta^\e\left(w/\bar w_L\right) \to \frac\mu{2\delta} \left( \left(\frac w{\bar w_L}\right)^\delta - 1 \right).
\ee
As explained in \cite{FPTT}, the motivation leading to the scaling of the frequency of interactions is clear. Since for $\e \ll1$  the microscopic interactions  produce a very small change of the value $w$, a finite  variation of the mean density can be observed only if agents in the system undergo a huge number of interactions  in a fixed period of time. 
By using the same scaling  one can easily observe a well-defined evolution equation for  the second moment of $f_\e(w,t)$, which will be well-defined also in the limit $\e \to 0$ (cf. the analysis in \cite{FPTT}).  
%%%%%%%%%%%%%%%%%%%%%%%%%%%%%%%%%%%%%%%%%%

The next step is to clarify why the linear kinetic model \fer{kin-w} modifies with $\e$ to produce in the limit $\e \to 0$ the Fokker--Planck type equation \fer{FPori} \cite{FPTT}. 
Given a smooth function $\varphi(w)$, and a collision of type \fer{coll} that produces a small variation of the difference $w_*-w$, using \fer{scal} and \fer{scal2} one obtains
\[
%\label{cor2}
\langle w_* -w \rangle = - \e \,A_{\delta,\e}(w)\,w;  \quad  \langle (w_* -w)^2\rangle =  \left(\e^2 \,  \,A_{\delta,\e}^2(w) + \e \lambda\right) w^2.
\]
Therefore, equating powers of $\e$,  it holds
\[
%\label{tay}
\langle \varphi(w_*) -\varphi(w) \rangle =  \e \left( - \varphi'(w)\, w\, \frac\mu{2\delta} \left( \left(\frac w{\bar w_L}\right)^\delta - 1 \right)
+ \frac \lambda 2 \, \varphi''(w)  w^2 \right) + R_\e (w),
\]
where the remainder term $R_\e(w)$, for a suitable $0\le \theta \le 1$ is such that 
\[
\frac 1\e \, R_\e(w) \to 0
\]
as  $\e \to 0$. Therefore, the time variation of the (smooth) observable quantity $\varphi(w)$  satisfies
\[
\begin{aligned}
%  \label{m-l23}
& \frac{d}{dt}\int_{\R_+}\varphi(w) \,f_\e(w,t)\,dw  = \\
& \int_{\R_+} \left( - \varphi'(w)\,w \,\frac\mu{2\delta} \left( \left(\frac w{\bar w_L}\right)^\delta - 1 \right)   + \frac \lambda 2 \varphi''(w) w^2 \right) f_\e(w,t)\, dw \ + \frac 1\e \mathcal R_\e(t) ,
\end{aligned}
\]
where $\mathcal R_\e(t)$ denotes the integral remainder term
\[
\label{rem3}
\mathcal R_\e(t) = \int_{\R_+ } R_\e(w)  f_\e(w,t)\, dw. 
\]
Letting $\e \to 0$,  shows that in consequence of the scalings \fer{scal} and \fer{scal2} the weak form of the kinetic model \fer{kin-w} is well approximated by the weak form of a linear Fokker--Planck equation (with variable coefficients)
\begin{equations}
	\label{m-13}
	& \frac{d}{dt}\int_{\R_+}\varphi(w) \,g(w,t)\,dw  = \\
	& \int_{\R_+} \int_{\R_+} \left( - \varphi'(w) \,w \,\frac\mu{2\delta} \left( \left(\frac w{\bar w_L}\right)^\delta - 1 \right)  + \frac \lambda 2 \varphi''(w) w^2 \right) g(w,t)\, dw.
\end{equations}  
In \fer{m-13} the density function $g(w,t)$ coincides with the limit, as $\e \to 0$, of the density $f_\e(w,t)$ \cite{FPTT}.   
Provided the boundary terms produced by the integration by parts vanish,  equation \fer{m-13} coincides with the weak form of the Fokker--Planck equation
\begin{equation}\label{FP2}
\frac{\partial g(w,t)}{\partial t} = \frac \lambda 2 \frac{\partial^2 }{\partial w^2}
\left(w^2 g(w,t)\right )+ \frac \mu{2}
\frac{\partial}{\partial w}\left( \left( \frac 1\delta\left(\frac w{\bar w_L}\right)^\delta - 1 \right) w \, g(w,t)\right).
\end{equation}
Equation \fer{FP2} describes the evolution of the distribution density $g(w,t)$ of the level of alcohol consumption $w \in \R_+$, in the limit of the \emph{grazing} interactions.  As often happens with Fokker-Planck type equations, the steady state density can be explicitly evaluated, and it results to be a generalized Gamma density, with parameters linked to the details of the microscopic interaction \fer{coll}.

 %%%%%%%%%%%%%%%%%%%%%%%%%%%%%%%%%%%%%%%%%%%%%%%%%%%%%%%%%%%%%%%%%%%
\subsection{Steady states are generalized Gamma densities}

%%%%%%%%%%%%%%%%%%%%%%%%%%%%%%%%%%%%%%%%%%%%%%%%%%%%%%%%%%%%%%%%%%%%%%%%%
Let us set $\gamma = \mu/\lambda$, and suppose that $\gamma>\delta$. In this case the stationary distribution of the Fokker--Planck equation \fer{FP2} is an integrable function which can be found explicitly by solving the first order differential equation 
\be\label{sd}
\frac{d }{dw}
\left(w^2 g(w)\right )+ 
\frac\gamma\delta\left(\left(\frac w{\bar w_L}\right)^\delta - 1 \right) w \, g(w)  =0.
\ee
Indeed, solving \fer{sd} with respect to $h(w)= w^2 g(w)$ by separation of variables allows to conclude that the unique solutions to \fer{sd}  are the functions 
\be\label{equilibrio}
g_\infty(w) =  g_\infty(\bar w_L)\left( \frac w{\bar w_L} \right)^{\gamma/\delta -2}  \exp\left\{ - \frac \gamma{\delta^2}\left( \left( \frac w{\bar w_L} \right)^\delta -1 \right)\right\}.
\ee 
If we fix the mass of the steady state \fer{equilibrio} equal to one,  the consequent probability density is a generalized Gamma. The generalized Gamma is characterized in terms of a shape $\kappa>0$, a scale parameter $\theta >0$, and an exponent $\delta >0$, and reads \cite{Lie,Sta}
\be\label{gg}
h(w,\kappa,\delta, \theta) = \frac\kappa{a^\kappa \Gamma \left( \kappa/\delta\right)}\, w^{\kappa -1} \exp\left\{ -\left( \frac w\theta\right)^\delta \right\}.
\ee
The shape and the scale paremeter of the equilibrium state \fer{equilibrio} are given by  
\be\label{para}
\kappa =  \frac\gamma\delta -1,  \quad  \theta = \bar w_L \left( \frac{\delta^2}\gamma\right)^{1/\delta}.
\ee
The limit $\delta \to 0$ in the Fokker--Planck equation \fer{FP2} corresponds to the drift term induced by the value function \fer{vf}. In this case, the equilibrium distribution \fer{equilibrio} takes the form of a Lognormal density \cite{GT17}
\be\label{l-n}
f_\infty(w) = \frac 1{\sqrt{2\pi \sigma}\, w} 
\exp\left\{ - \frac{(\log w - m)^2}{2 \sigma}\right\},
\ee 
where 
\be\label{pa}
\sigma = \frac 1\gamma,  \quad m = \log \bar w_L - \sigma.
\ee 
Note that for all values $\delta >0$ the moments are expressed in terms of the parameters $\bar w_L$, $ \delta$, $\lambda$, and $\mu$, denoting respectively the limit level of alcohol consumption, the level of addiction the variance $\lambda$ of the random effects and the values $ \delta$ and $\mu$ characterizing  the value function $\phi_\delta^\e$ defined in \fer{vd}. The condition $\gamma >\delta$, that guarantees the existence of a steady state density,  is achieved for small values of $\delta$, or for sufficiently large values of $\gamma$. Since $\gamma = \mu/ \lambda$,  the quotient between the maximal amount of level of change of alcohol consumption and the variance of the random change, the condition is satisfied when the random effects are very small. 

It is interesting to remark that, in the case of the gamma distribution, the mean value of the equilibrium density \fer{equilibrio} is always less than $\bar w_L$. In this case in fact
\[
M_1 = \kappa\,\theta = \bar w_L \frac{\gamma -1}\gamma. 
\]

\begin{remark} {\rm The analysis of the possible variety of steady states densities generated by the value functions \fer{vd} allows to conclude that the best fitting of the level of alcohol consumption could be obtained by a generalized Gamma distribution. Indeed, the three cases treated in \cite{Keh,Reh} are included as particular cases of generalized Gamma distributions. The data in \cite{Keh} enlighten the fact that, in the Weibull case, the characteristic exponent $\delta \ge 1/2$. This helps to understand why the Lognormal distribution, which is obtained by pushing $\delta$ to zero, probably fails to give a good fitting of the level of alcohol consumption. It is remarkable that in \cite{Par}, by noticing that the generalized Gamma nests the Gamma, Weibull and Lognormal distributions as special cases, it has been proposed as useful test distribution for the problem under investigation, since, if all other distributions are rejected, the generalized Gamma results minimally restrictive and less susceptible to misspecification bias \cite{Man}.}
\end{remark}

\begin{remark} {\rm Once the shape of the alcohol consumption distribution has been characterized, the next step would be to find the parameters of the generalized Gamma that provide the best possible fit of  real datasets. However, the parameter estimation of the generalized Gamma distribution  is a very difficult problem.  This
		topic was dealt with in many papers (cf. the results in \cite{Bas,Gom,Man,Yil} and the list of references therein) but since almost all of the proposed methods are quite complex, this topic is still an open area, that needs to be better studied.}
\end{remark}

%%%%%%%%%%%%%%%%%%%%%%%%%%%%%%%%%%%%%%%%%%%%%%%%%%%%%%%%%%
\section{Numerical experiments}\label{numerics}
%%%%%%%%%%%%%%%%%%
The main interest related to the Fokker--Planck equation \fer{FP2} is the analytical description of the equilibrium, that in the present case is given by the generalized Gamma distribution \fer{equilibrio}. Except in some simple cases \cite{BaTo,BaTo2}, kinetic models of Boltzmann type do not allow similar results. On the other hand, the kinetic equation \fer{kin-w} is the right-one to study, since it contains all the microscopic details that are apparently lost in the \emph{grazing} limit procedure.
For this reason, we report in this Section several numerical experiments that describe  convergence to the steady state (identified by the generalized Gamma distribution  in the limit $\e \to 0$ in the Boltzmann dynamics (\ref{coll}) for various values of the parameter $\e$. In details, our study proceeds as follows. We consider two different values for the two parameters $\gamma$ and $\bar w_L$ of the model. The first parameter fixes the ratio between the intensity of the value function $\Phi_\delta^\epsilon$ and the variance of a random variation in the behavior of the agents. In  the Boltzmann dynamics, we consider a uniform random variable  with zero mean and variance $1/12$. The second parameter $\bar w_L$  measures the threshold value that it would be better not to exceed to enter into alcohol addiction. The values considered in the experiments are $[\gamma,\bar w_L]=[1,6]$ and $[\gamma,\bar w_L]=[2,3]$.

For these two set of parameters, we study the behavior of the microscopic dynamics detailed in (\ref{coll}) for different values of the parameter $\delta$ in the value function. This parameter is of paramount importance, since it modifies the shape of the steady states in the Fokker--Planck description. Depending on its value, from the generalized Gamma distribution we obtain a Gamma density ($\delta=1$), a Weibull density ($\delta(1+\delta)=\gamma$) or a Lognormal distribution ($\delta\rightarrow 0$) as equilibrium states. 

\begin{table}[ht]
		%\vspace{+0.3cm}
	\scriptsize
	\caption{Expectation and variance of the Fokker-Planck and of the Boltzmann model for different values of the scaling factor $\epsilon$ and the shape factor $\delta$ for the Test 1.}\label{table1}
	\setlength\extrarowheight{5pt}
	\centering
	\begin{tabular}{|p{1cm}|p{2cm}|p{2.5cm}|p{2.5cm}|  }
		\hline
		\multicolumn{2}{|c|}{$\gamma =1$, $\bar w_L =6$} & \multicolumn{1}{c|}{$E[g_\infty^B(w)]$} & \multicolumn{1}{c|}{$Var(g_\infty^B(w))$}\\
		\hline
		\multirow{5}{*}{{$\delta=0.1$}}
		& $\epsilon=0.9$& 2.66 
		& 24.05
		\\
		\cline{2-4}
		&$\epsilon=0.5$& 3.03     &  18.51
		\\
		\cline{2-4}
		&$\epsilon=0.1$ &3.32      
		&13.56\\
		\cline{2-4}
		& $\epsilon=0.001$ & 3.36    
		& 18.21 \\
		\cline{2-4}
		& FP & 3.37    
		& 18.22 \\
		\hline
		\hline
		\multirow{5}{*}{{$\delta=0.3$}} 
		& $\epsilon=0.9$& 2.19
		& 19.58
		\\
		\cline{2-4}
		&$\epsilon=0.5$&  2.52   &  14.93
		\\
		\cline{2-4}
		&$\epsilon=0.1$ &   2.81  
		&14.93\\
		\cline{2-4}
		& $\epsilon=0.001$ &  2.85   
		& 15.05\\
		\cline{2-4}
		& FP &  2.85   
		& 14.93\\
		\hline
		\hline
		\multirow{5}{*}{{$\delta=0.5$}} 
		& $\epsilon=0.9$& 1.60
		& 14.45
		\\
		\cline{2-4}
		&$\epsilon=0.5$&  1.93   &  11.31
		\\
		\cline{2-4}
		&$\epsilon=0.1$ &   2.20  
		&11.52\\
		\cline{2-4}
		& $\epsilon=0.001$ &    2.23 
		&11.61 \\
		\cline{2-4}
		& FP &    2.25 
		&11.65 \\
		\hline
		\hline
		\multirow{5}{*}{{$\delta=0.8$}} 
		& $\epsilon=0.9$& 0.38
		& 3.57
		\\
		\cline{2-4}
		&$\epsilon=0.5$& 0.77    &4.70  
		\\
		\cline{2-4}
		&$\epsilon=0.1$ &     1.02
		&5.64\\
		\cline{2-4}
		& $\epsilon=0.001$ &  1.06   
		&5.80 \\
		\cline{2-4}
		& FP &  1.14   
		&6.17 \\
		\hline
	\end{tabular}

\end{table}

\begin{table}[ht]
	%	\vspace{+0.3cm}
	\scriptsize
	\caption{Expectation and variance of the Fokker-Planck and of the Boltzmann model for different values of the scaling factor $\epsilon$ and the shape factor $\delta$ for the Test 2.}\label{table2}
	\setlength\extrarowheight{5pt}
	\centering
		\begin{tabular}{|p{1cm}|p{2cm}|p{2.5cm}|p{2.5cm}|  }
		\hline
		\multicolumn{2}{|c|}{$\gamma =2$, $\bar w_L =3$} & \multicolumn{1}{c|}{ $E[g_\infty^B(w)]$} & \multicolumn{1}{c|}{ $Var(g_\infty^B(w))$}\\
		\hline
		\multirow{5}{*}{{$\delta=0.1$}} 
		& $\epsilon=0.9$& 2.27
		& 6.92
		\\
		\cline{2-4}
		&$\epsilon=0.5$&2.13     &  3.77
		\\
		\cline{2-4}
		&$\epsilon=0.1$ &     2.24
		&3.37\\
		\cline{2-4}
		& $\epsilon=0.001$ &     2.25
		& 3.27\\
		\cline{2-4}
		&FP &     2.25
		& 3.09\\
		\hline
		\hline
		\multirow{5}{*}{{$\delta=0.3$}} 
		& $\epsilon=0.9$& 2.10 
		&6.09 
		\\
		\cline{2-4}
		&$\epsilon=0.5$& 1.98     & 3.37 
		\\
		\cline{2-4}
		&$\epsilon=0.1$ & 2.09    
		&3.05\\
		\cline{2-4}
		& $\epsilon=0.001$ & 2.11     
		& 3.01\\
		\cline{2-4}
		& FP & 2.11     
		& 2.89\\
		\hline
		\hline
		\multirow{5}{*}{{$\delta=0.5$}} 
		& $\epsilon=0.9$& 1.91 
		& 5.34
		\\
		\cline{2-4}
		&$\epsilon=0.5$& 1.82     &  3.04
		\\
		\cline{2-4}
		&$\epsilon=0.1$ & 1.94    
		&2.78\\
		\cline{2-4}
		& $\epsilon=0.001$ & 1.96     
		& 2.79\\
		\cline{2-4}
		&FP & 1.96     
		& 2.70\\
		\hline
		\hline
		\multirow{5}{*}{{$\delta=0.8$}} 
		& $\epsilon=0.9$& 1.60 
		& 4.33
		\\
		\cline{2-4}
		&$\epsilon=0.5$& 1.54     &  2.59
		\\
		\cline{2-4}
		&$\epsilon=0.1$ &  1.67   
		&2.44\\
		\cline{2-4}
		& $\epsilon=0.001$ & 1.69     
		& 2.41\\
		\cline{2-4}
		& FP & 1.70     
		& 2.43\\
		\hline
	\end{tabular}
\end{table}

\begin{figure}\centering
	{\includegraphics[width=6.5cm]{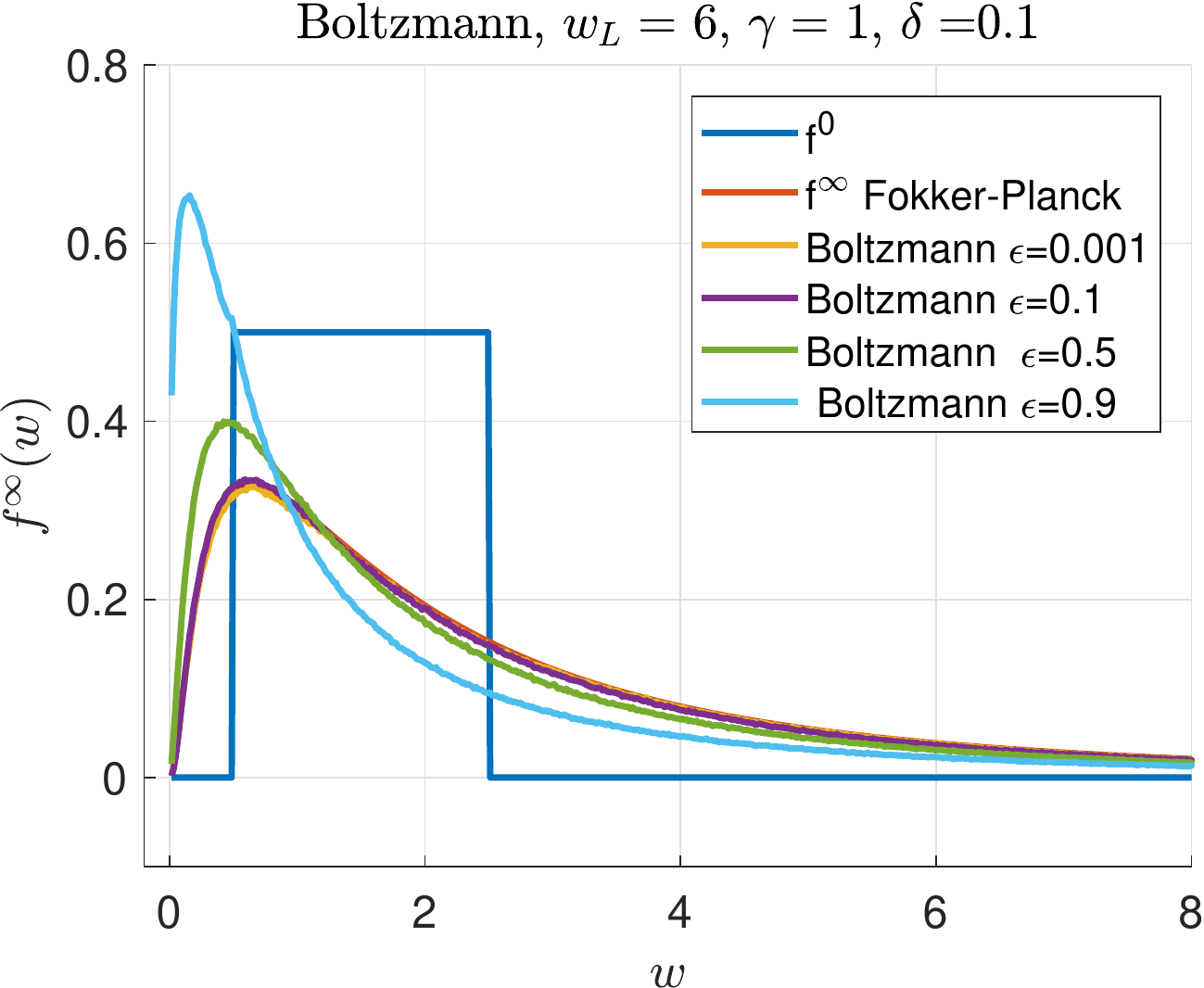}}
	\hspace{+0.35cm}
	{\includegraphics[width=6.5cm]{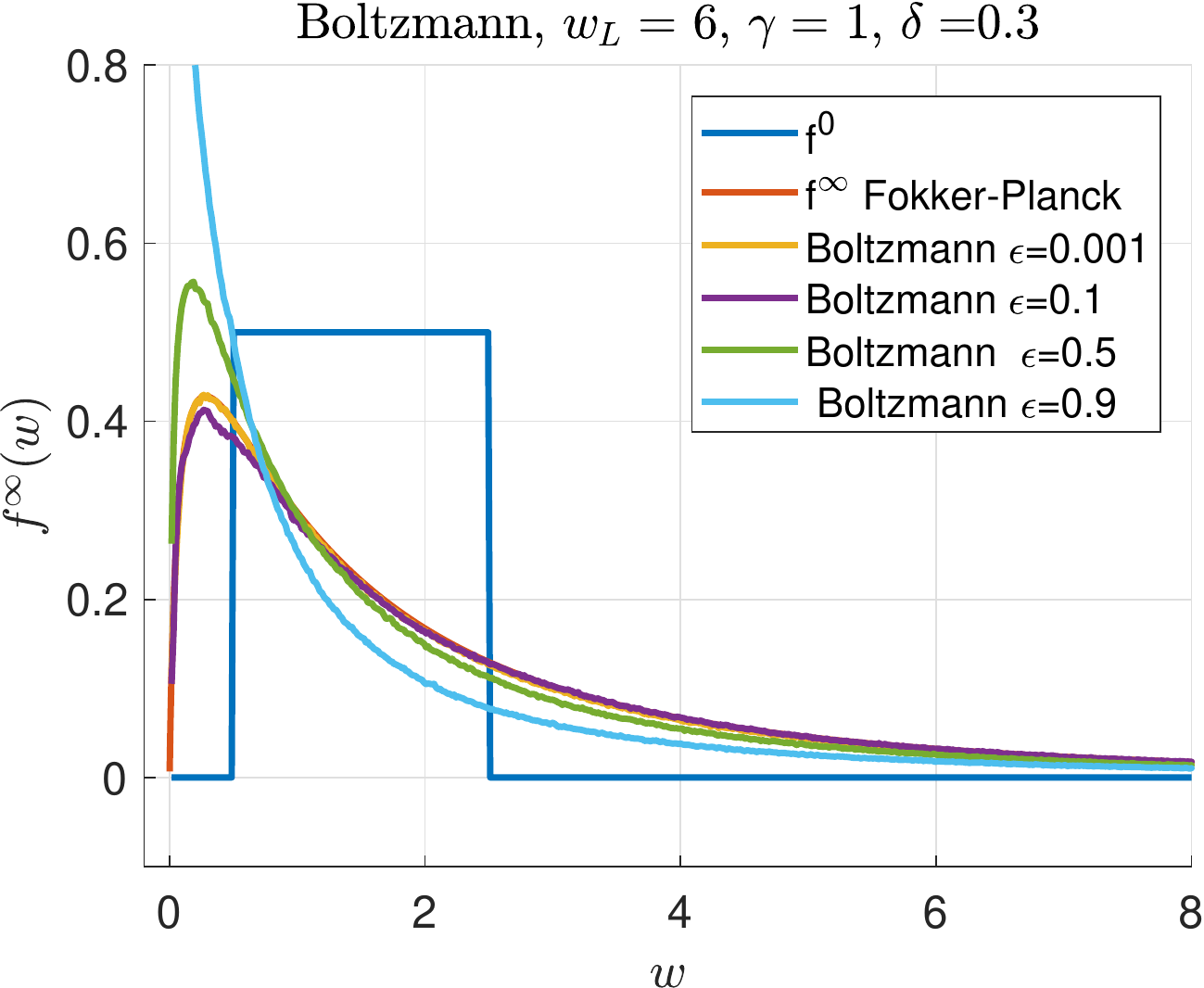}}\\
	\vspace{+0.45cm}
	{\includegraphics[width=6.5cm]{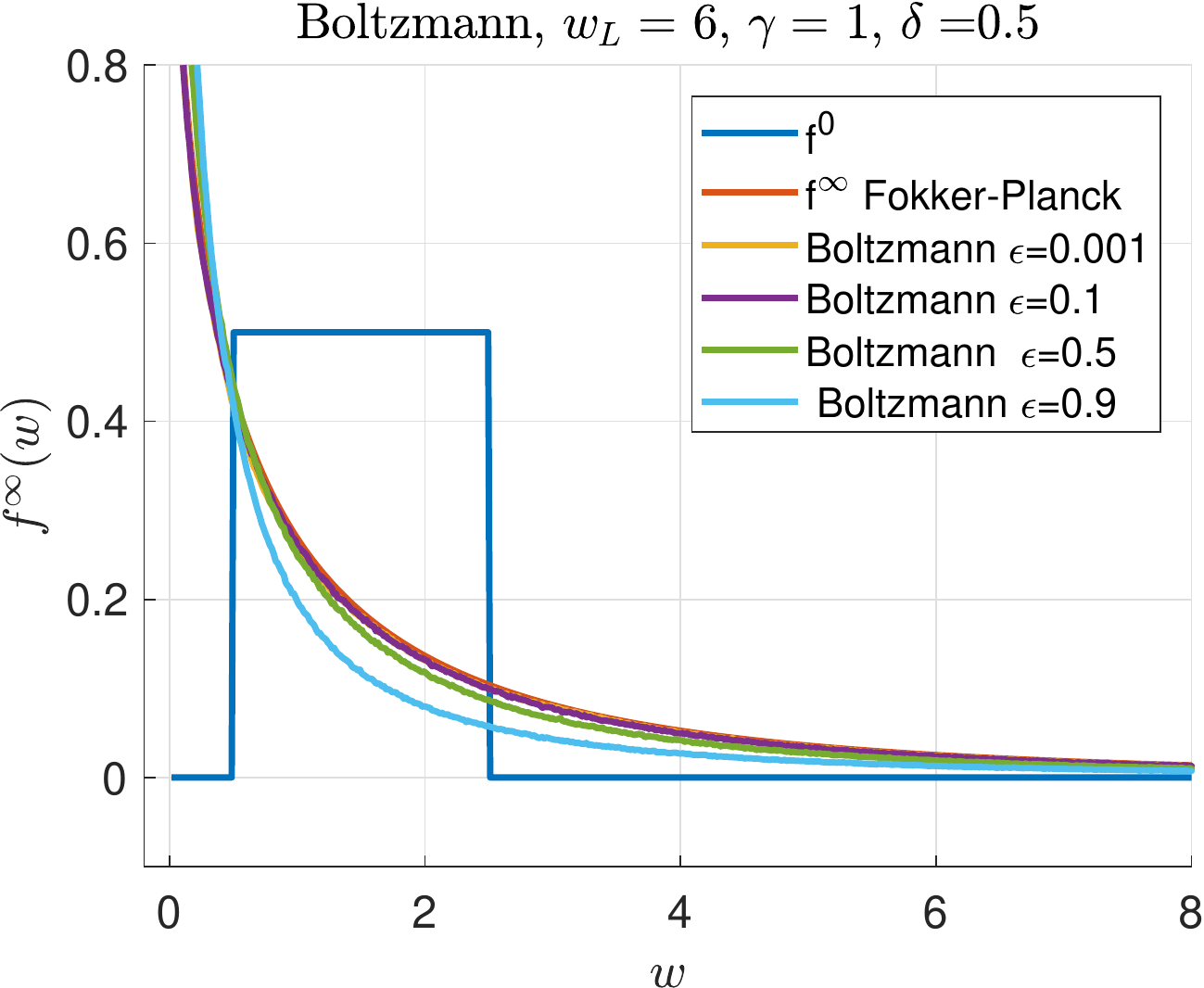}}
	\vspace{+0.35cm}
	{\includegraphics[width=6.5cm]{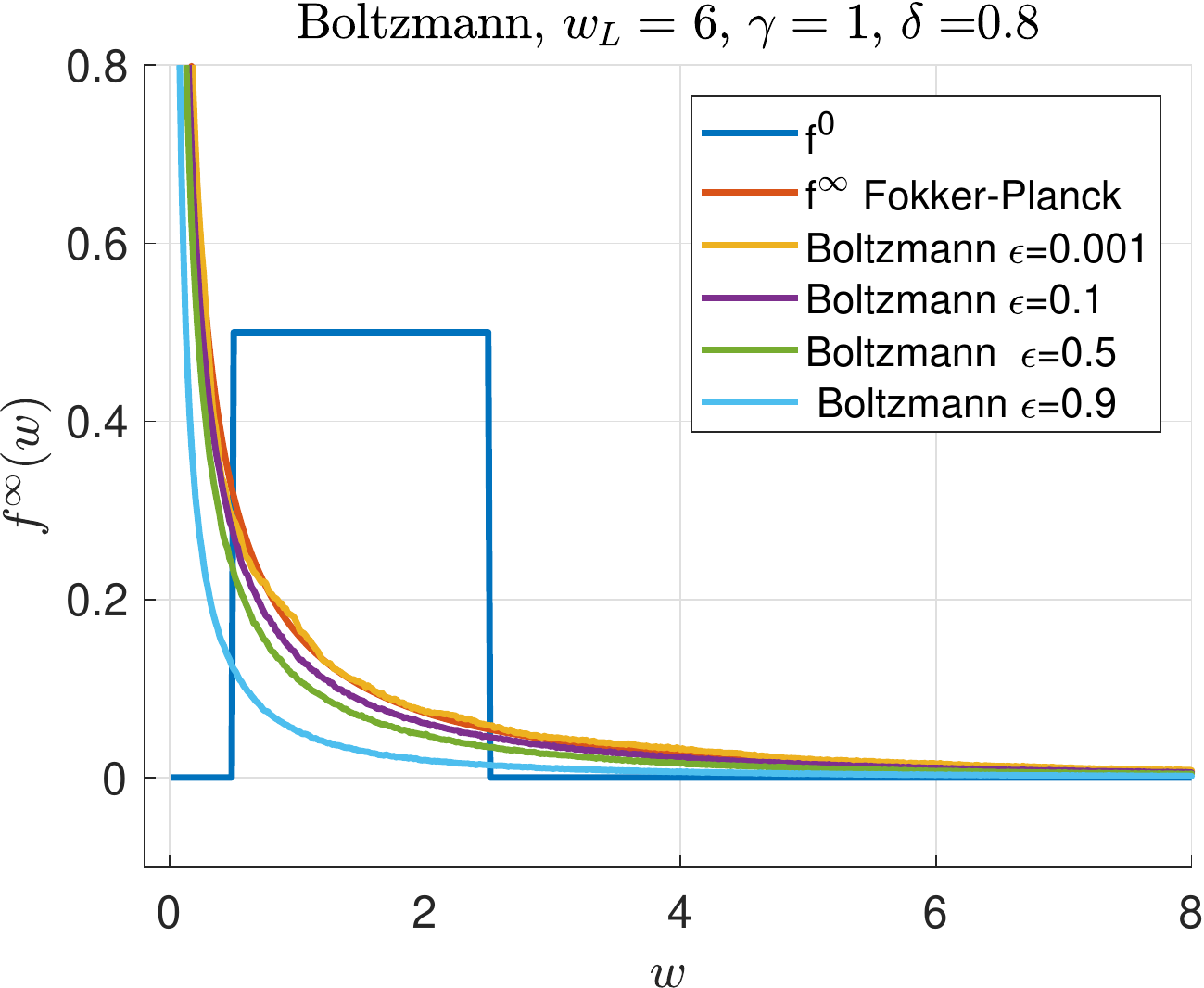}}
	\caption{Test 1. Convergence to the Fokker-Planck dynamics for the Boltzmann model as a function of the scaling parameter $\epsilon$. Top left $\delta=0.1$, top right $\delta=0.3$, bottom left $\delta=0.5$, bottom right $\delta=0.8$. }\label{fig:test1}
\end{figure}

\begin{figure}\centering
	{\includegraphics[width=6.5cm]{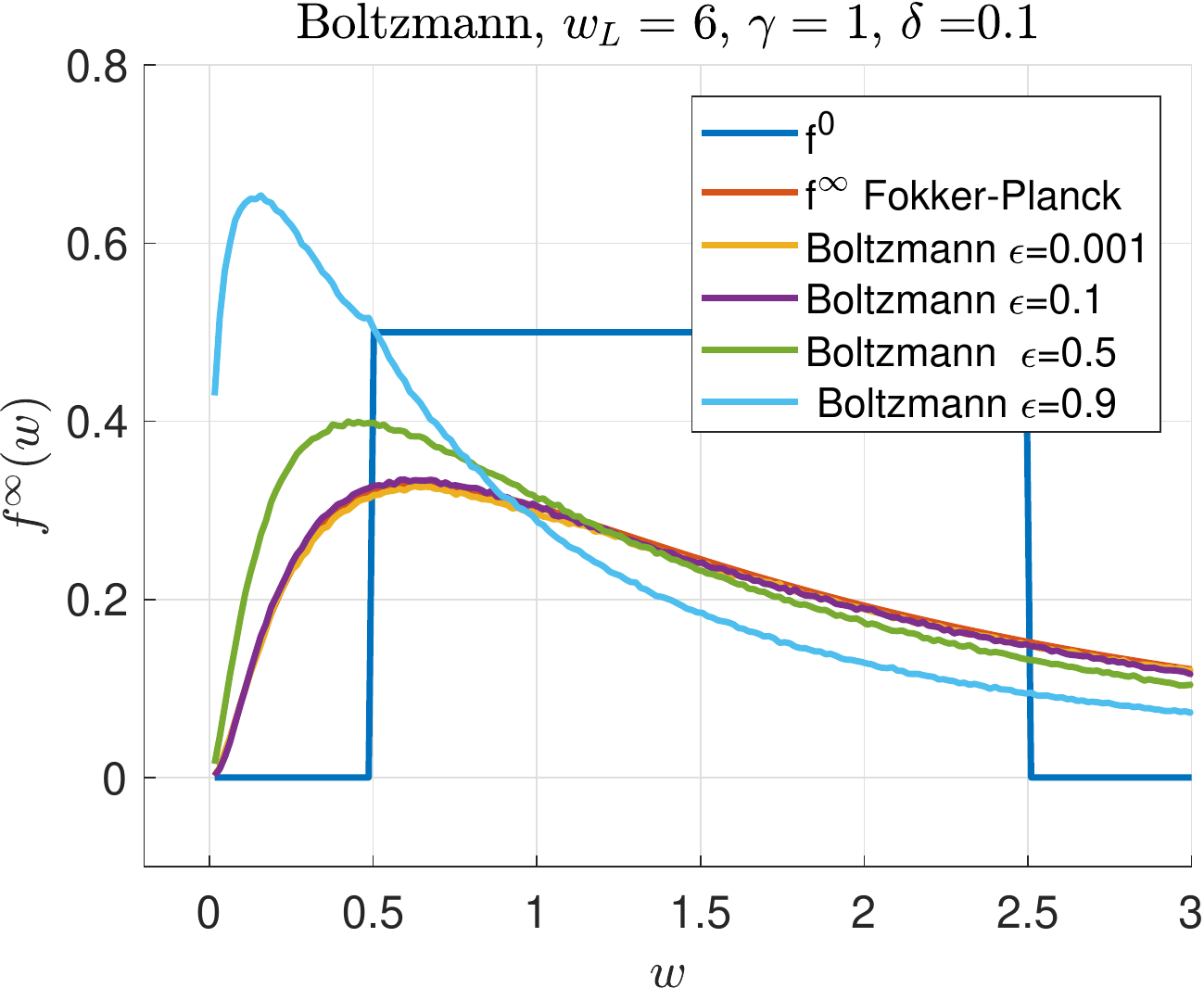}}
	\hspace{+0.35cm}
	{\includegraphics[width=6.5cm]{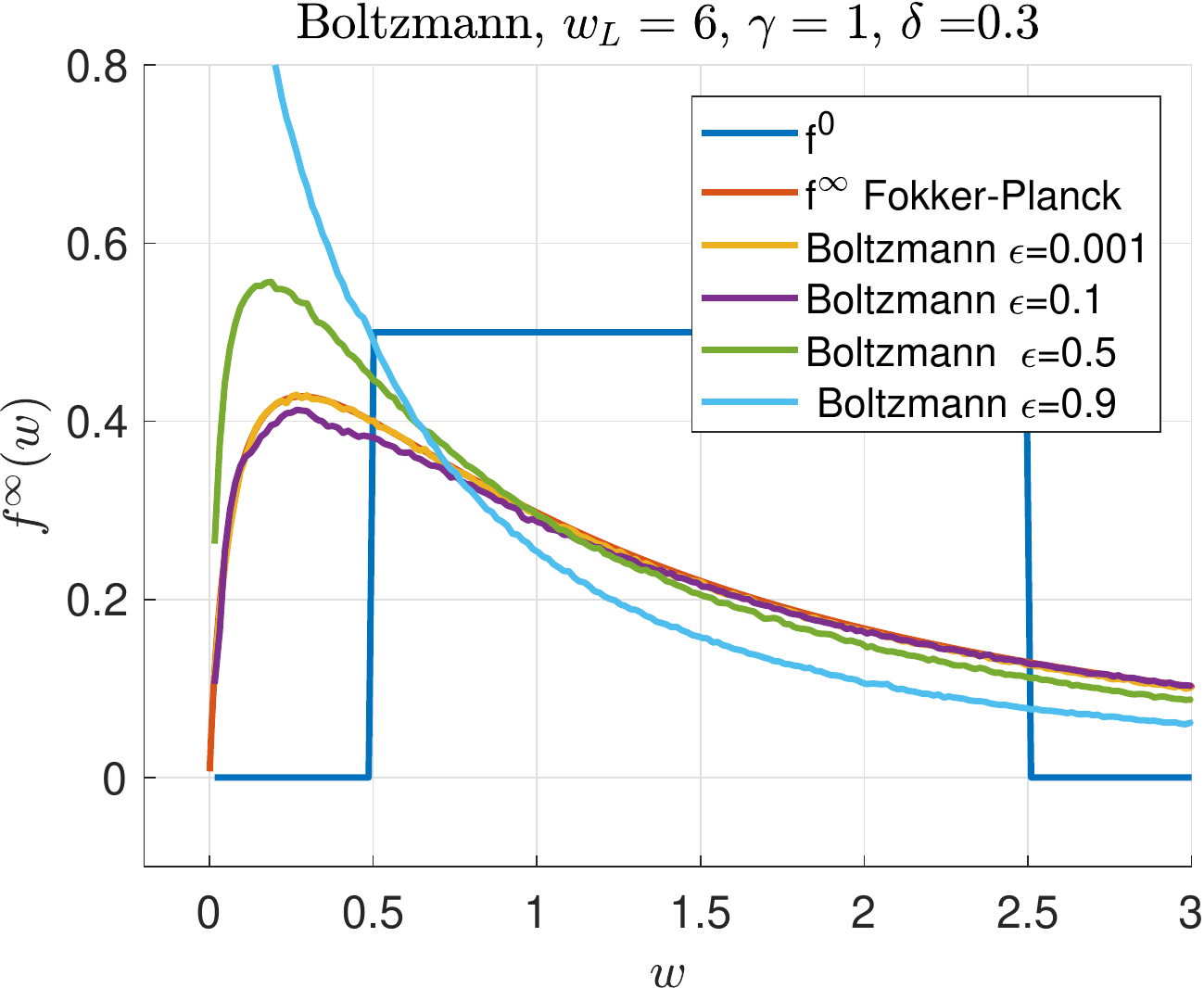}}\\
	\vspace{+0.45cm}
	{\includegraphics[width=6.5cm]{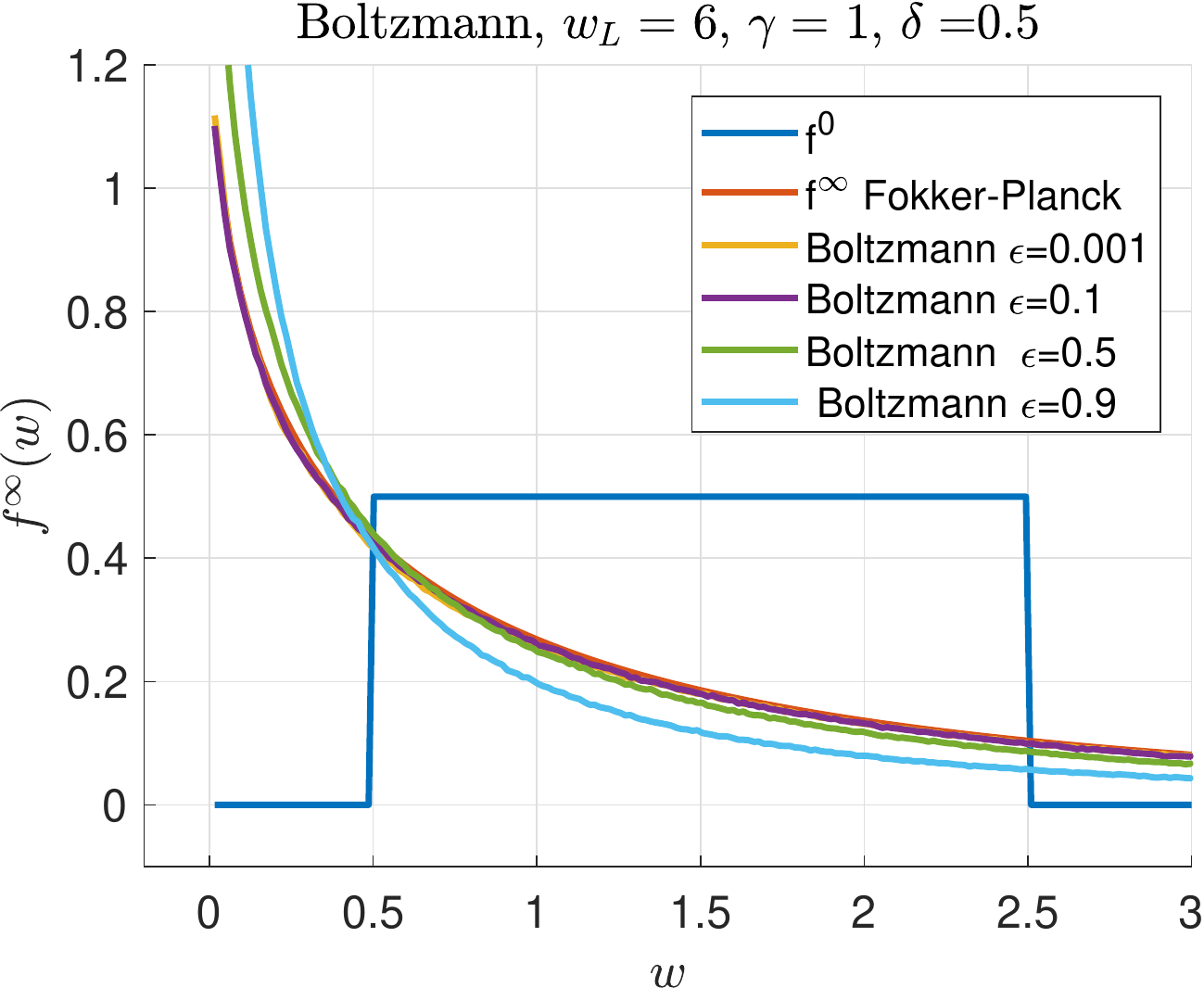}}
	\vspace{+0.35cm}
	{\includegraphics[width=6.5cm]{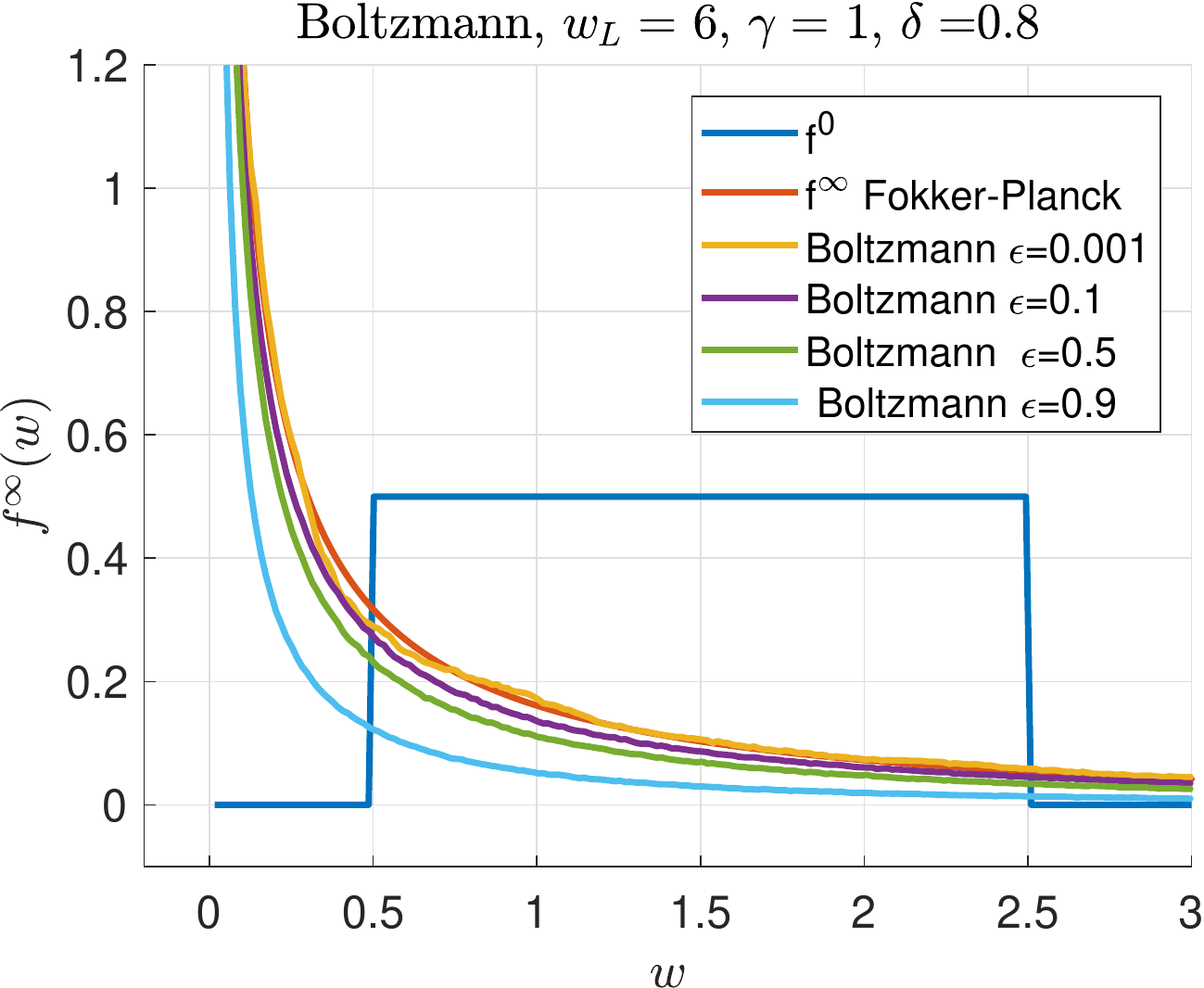}}
	\caption{Test 1. Convergence to the Fokker-Planck dynamics for the Boltzmann model as a function of the scaling parameter $\epsilon$. Top left $\delta=0.1$, top right $\delta=0.3$, bottom left $\delta=0.5$, bottom right $\delta=0.8$. Magnification of the solution around the origin. }\label{fig:test1_zoom}
\end{figure}

\begin{figure}\centering
	{\includegraphics[width=6.5cm]{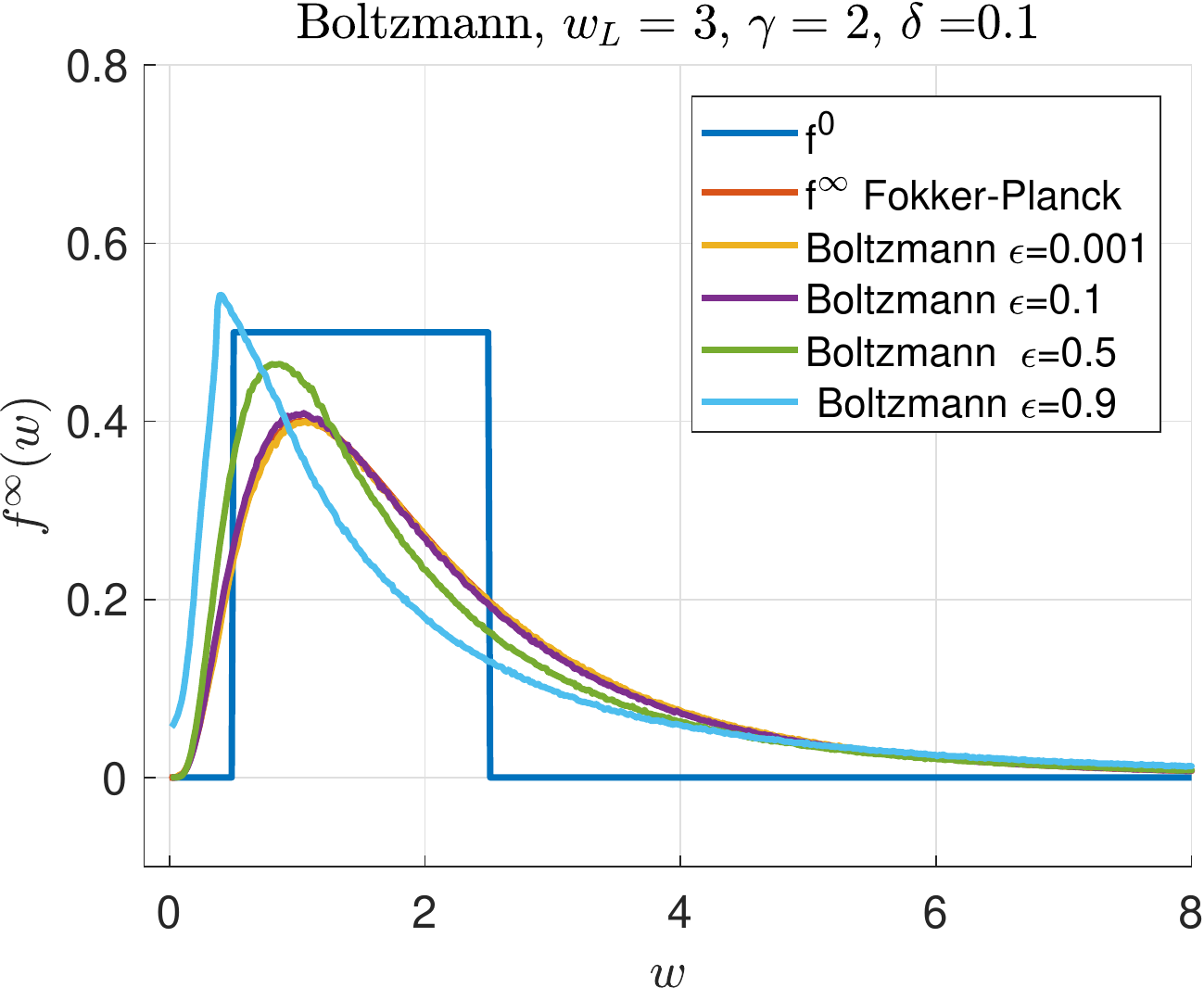}}
	\hspace{+0.35cm}
	{\includegraphics[width=6.5cm]{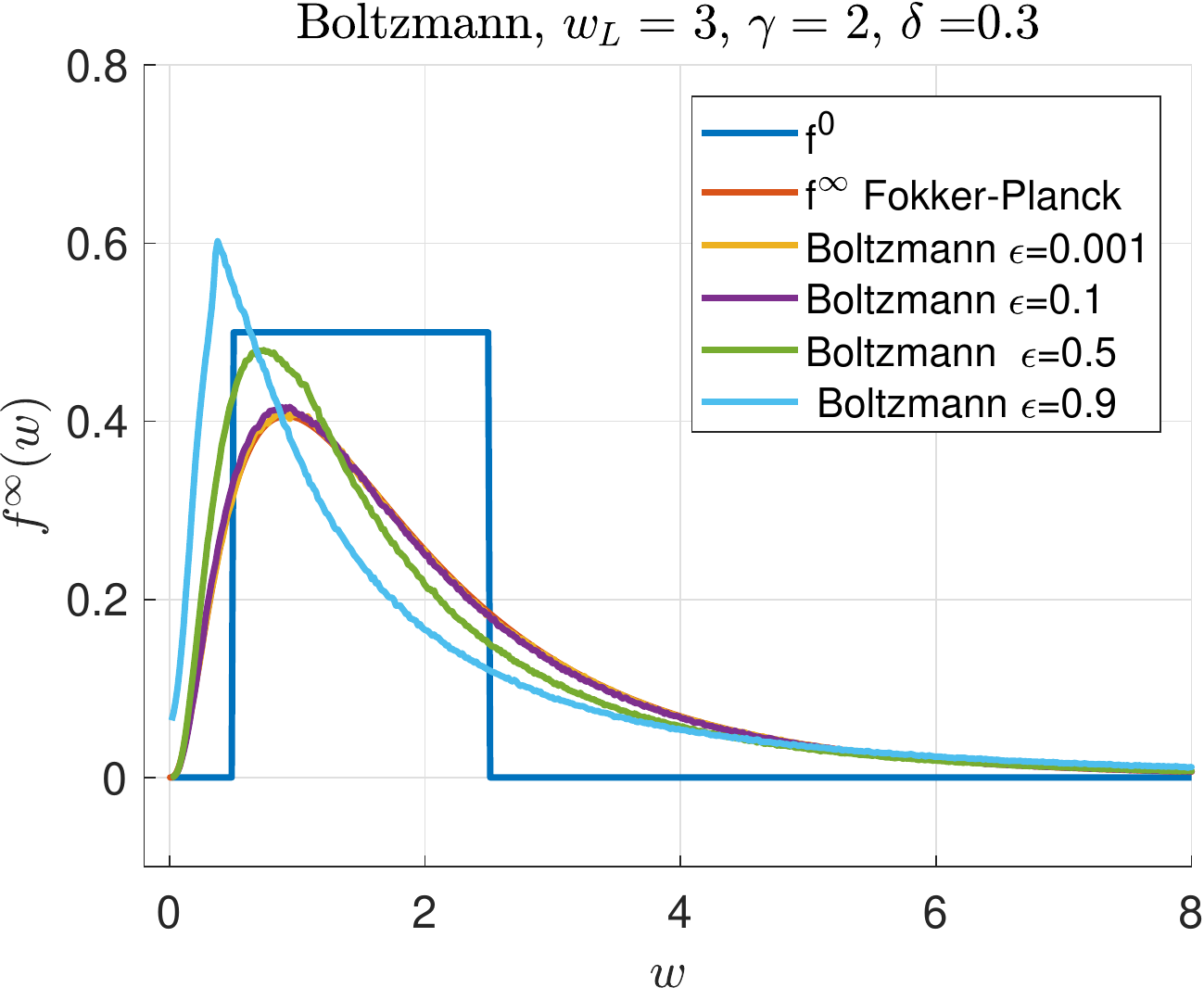}}\\
	\vspace{+0.45cm}
	{\includegraphics[width=6.5cm]{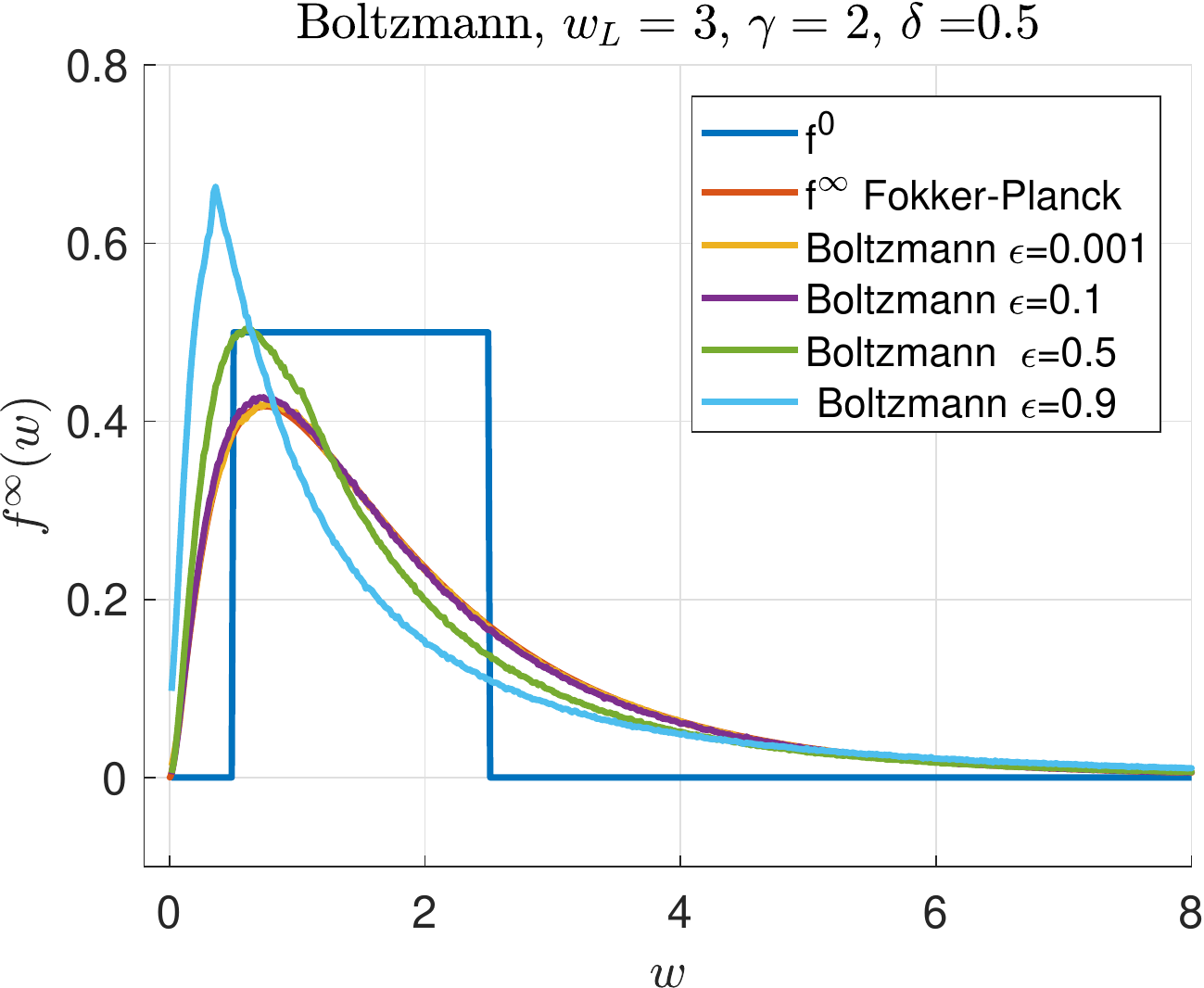}}
	\vspace{+0.35cm}
	{\includegraphics[width=6.5cm]{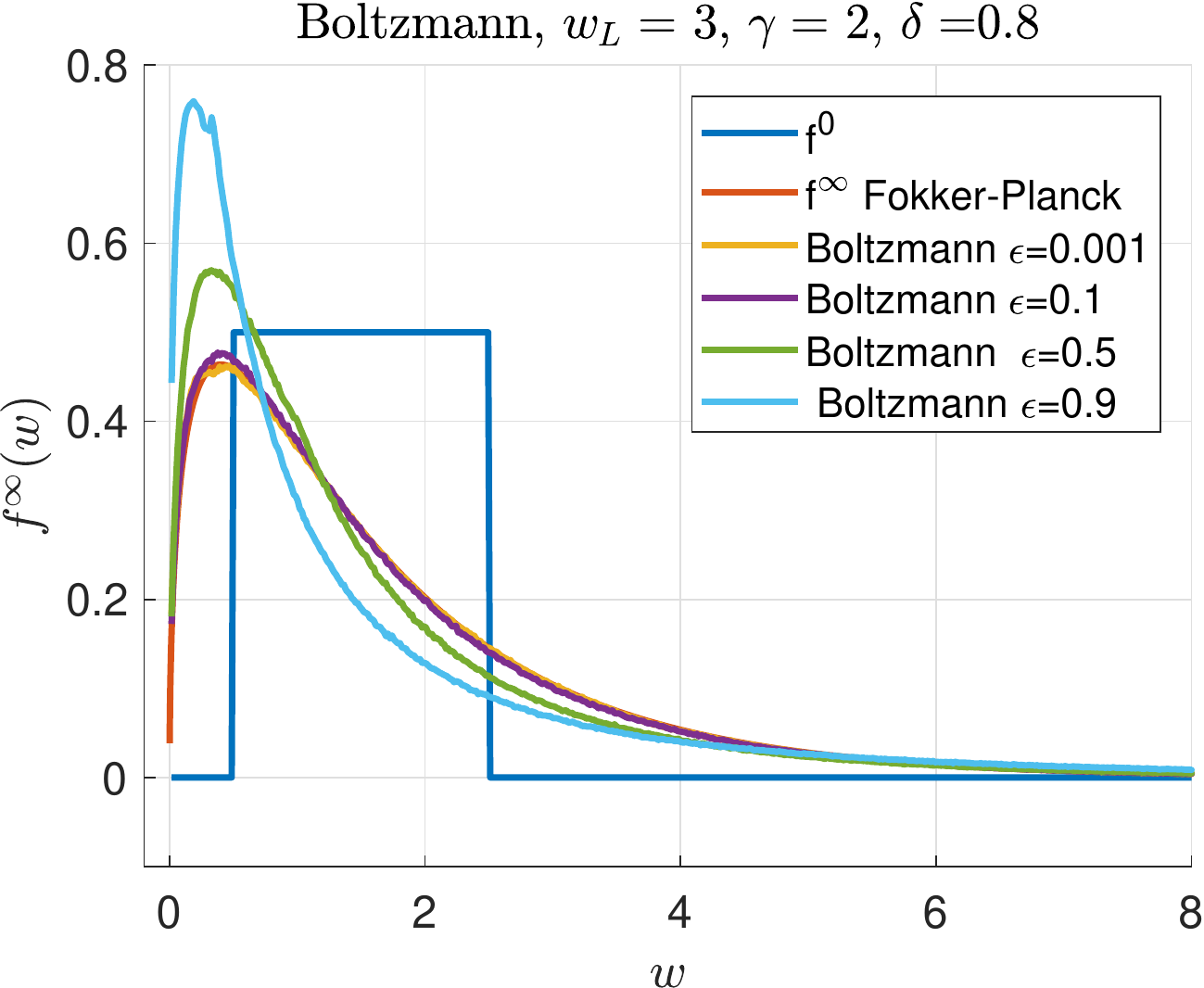}}
	\caption{Test 2. Convergence to the Fokker-Planck dynamics for the Boltzmann model as a function of the scaling parameter $\epsilon$. Top left $\delta=0.1$, top right $\delta=0.3$, bottom left $\delta=0.5$, bottom right $\delta=0.8$. }\label{fig:test2}
\end{figure}

\begin{figure}\centering
	{\includegraphics[width=6.5cm]{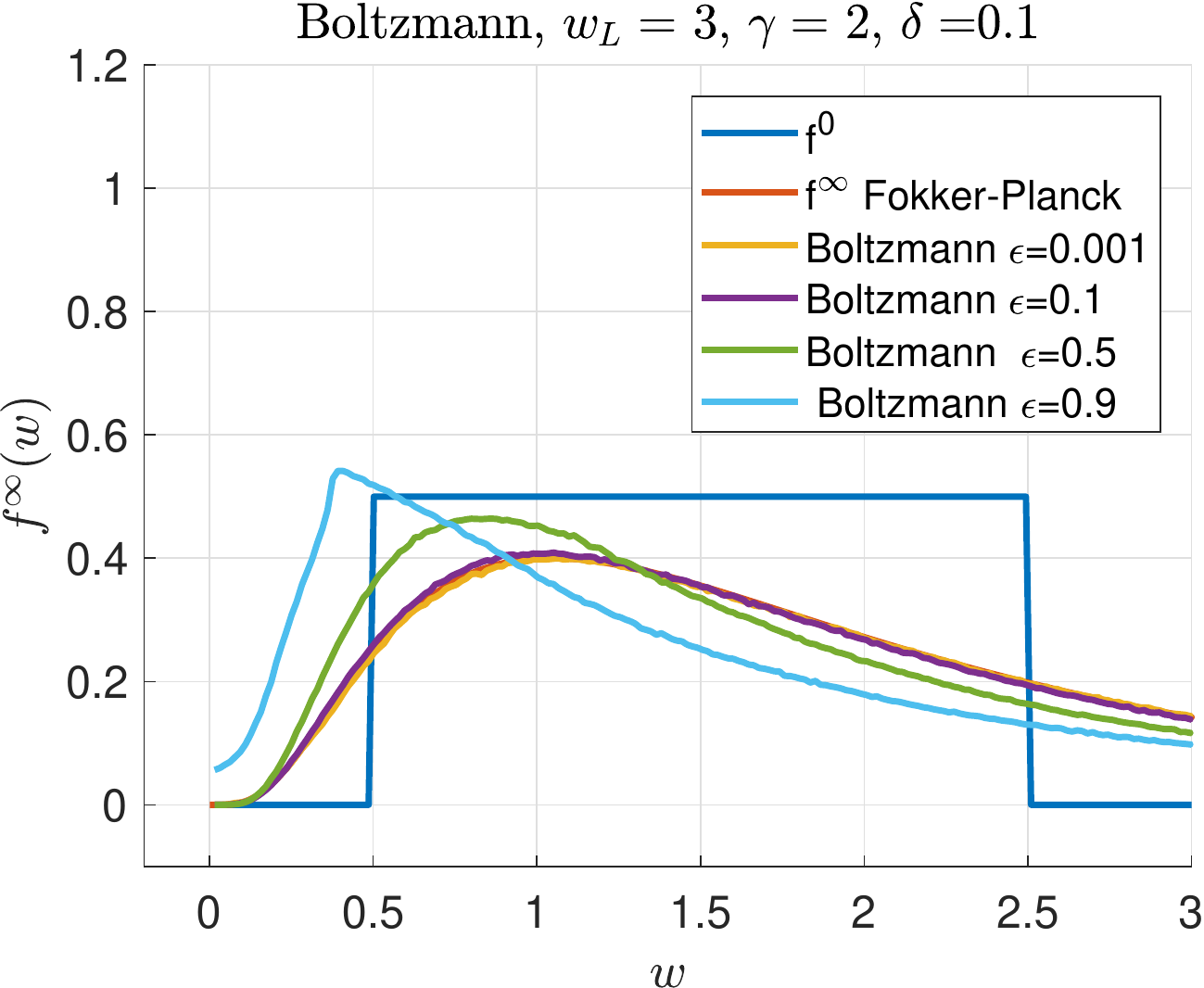}}
	\hspace{+0.35cm}
	{\includegraphics[width=6.5cm]{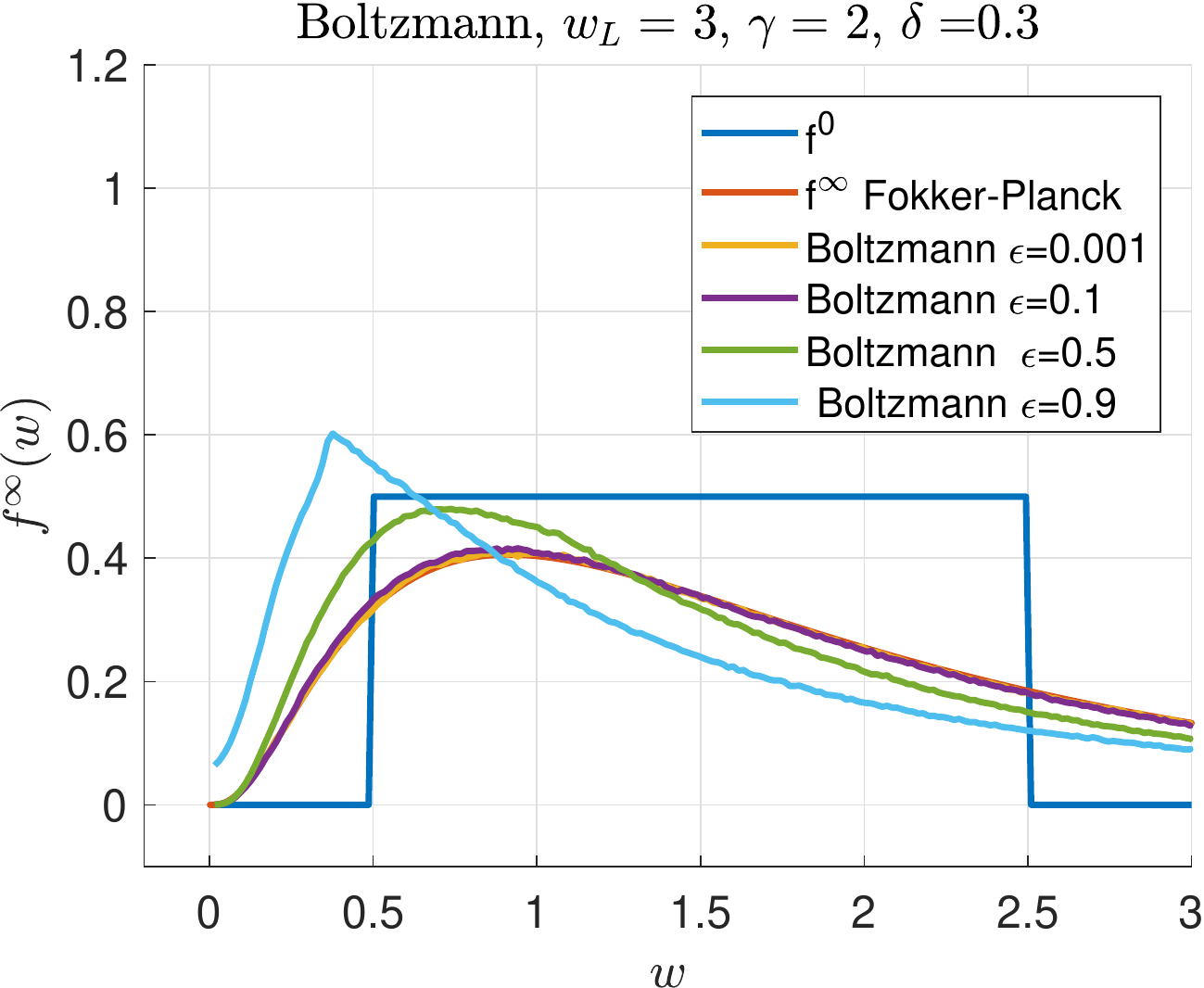}}\\
	\vspace{+0.45cm}
	{\includegraphics[width=6.5cm]{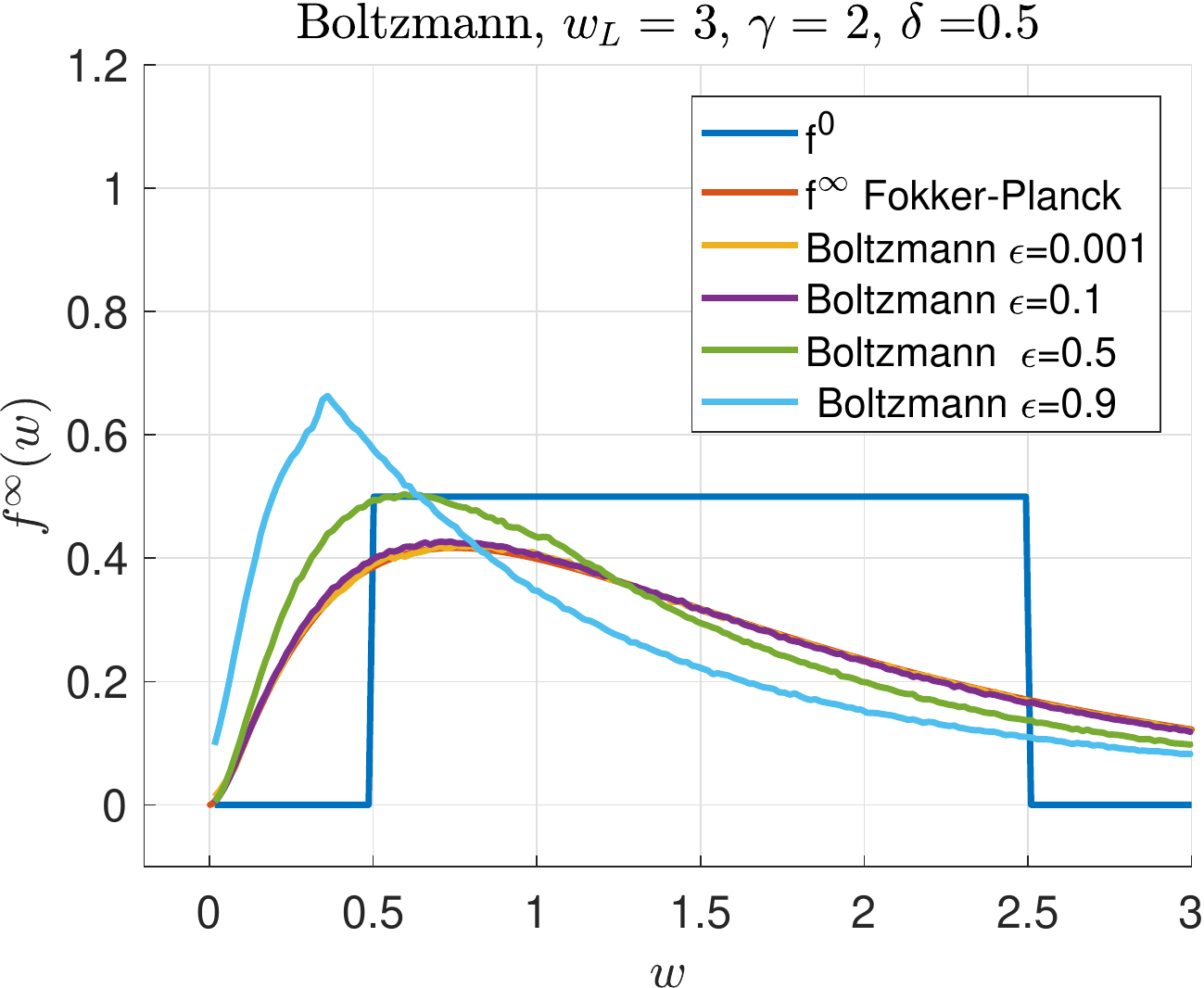}}
	\vspace{+0.35cm}
	{\includegraphics[width=6.5cm]{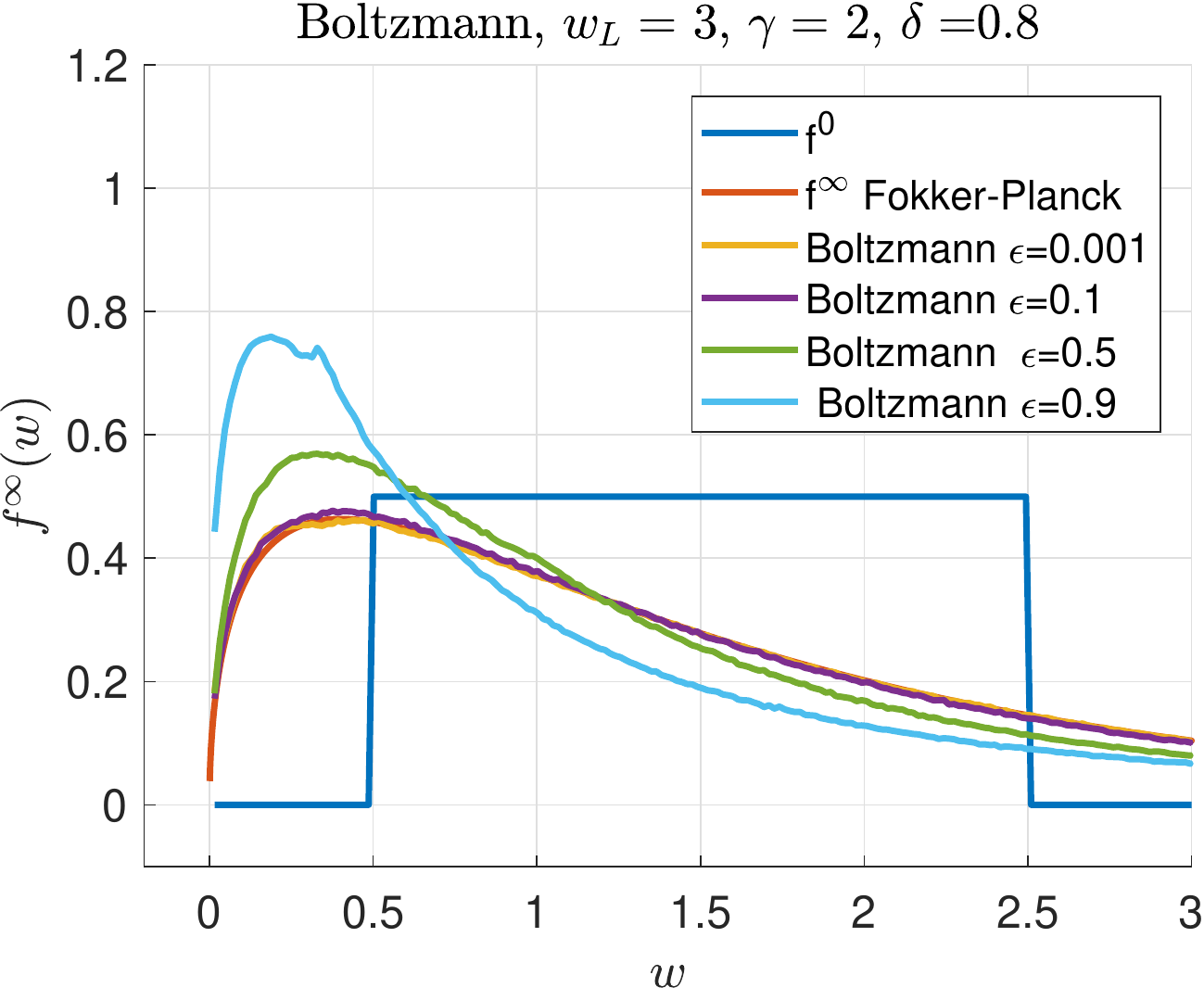}}
	\caption{Test 2. Convergence to the Fokker-Planck dynamics for the Boltzmann model as a function of the scaling parameter $\epsilon$. Top left $\delta=0.1$, top right $\delta=0.3$, bottom left $\delta=0.5$, bottom right $\delta=0.8$. Magnification of the solution around the origin. }\label{fig:test2_zoom}
\end{figure}

In order to show that the Fokker-Planck dynamics can be seen as a particular case of the Boltzmann one, we place ourselves in the same scaling in which the Fokker-Planck equation (\ref{FP2}) has been analytically derived. For this reason, we consider in the value function \fer{vd} a decreasing sequence of values for the scaling factor $\epsilon$ ranging from $0.9$ to $10^{-3}$. Correspondingly we scale the effects of the random part, according to \fer{scal}, and the interaction frequency according to \fer{scal2} to have an $\e$-free evolution of the mean. In Figure \ref{fig:test1} we show the different equilibrium states obtained with the Boltzmann dynamics for the different values of the scaling factor $\epsilon$. The Figure \ref{fig:test1_zoom} reports the same results of Figure \ref{fig:test1} with magnification around the origin to better appreciate the differences between the solutions.  Figure \ref{fig:test2} and \ref{fig:test2_zoom} show the same dynamics where $\gamma$ and $\bar w_L$ have been changed to respectively $2$ and $3$, namely by diminishing the threshold and  increasing the variability.

The curves have been obtained by using $10^3$ agents and by averaging the steady state solution over $10^5$ realizations.
In each image it is also reported the steady state of the Fokker-Planck model (\ref{equilibrio}) and the initial data, which are the same for both models. The initial value has been considered uniformly distributed in $[0.5,2.5]$ for all different situations analyzed. From top left to bottom right the shape factor $\delta$ takes the values $0.1$, $0.3$, $0.5$ and $0.8$ which causes the equilibrium state to completely change its shape. For all tested situations an extremely good agreement between the Boltzmann and the Fokker--Planck models is obtained from approximately $\epsilon=0.001$. However, very similar steady state profiles are found from $\e = 0.1$, which fully describes a kinetic evolution of Boltzmann type. This interesting behavior can be observed also from the results collected in the Tables \ref{table1} and \ref{table2}, which report the expected value and the variance obtained with the kinetic Boltzmann model for the different values of the parameter $\epsilon$ and the shape factor $\delta$.   

%%%%%%%%%%%%%%%%%%%%%%
\section{Appendix}\label{app}
In this short appendix we verify that, for any given constant $0 < \delta <1$ and $\e >0$, the value function $\Phi_\delta^\e(s)$ satisfies, for any given $s$ such that $0< s <1$ properties \fer{ccd} and \fer{cce}. Indeed, substituting into \fer{cce}  the value function \fer{vd} shows that property holds provided
\be\label{ok1}
\exp\left\{\frac \e\delta\left[(1+s)^\delta +(1-s)^\delta -2\right] \right\} \le 1.
\ee
Inequality \fer{ok1} holds since, for any given $0<s<1$, the function
\[
Z_s(\delta) = (1+s)^\delta +(1-s)^\delta -2 
\] 
is a convex function of $\delta$ that is equal to zero at the boundaries of the interval $0\le\delta\le 1$, and this clearly implies $Z_s(\delta) \le 0$ on this interval, independently of the value of $s$. 

Property \fer{ccd} holds since the value function $\Phi_\delta^\e(s)$ is concave with respect to $s \ge0$. Direct computations allow to conclude that the sign of the second derivative of the value function is non positive if
\be\label{ok2}
\exp\left\{ \frac\e\delta\left(s^\delta -1 \right) \right\} \ge \frac{\e s^\delta -(1-\delta)}{\e s^\delta +(1-\delta)}.
\ee
Inequality \fer{ok2} holds considering that, for $x \ge 0$, the function
\[
Y(x) =   \exp\left\{ \frac\e\delta\left(x -1 \right) \right\} - \frac{\e x -(1-\delta)}{\e x +(1-\delta)}
\]
is a convex function, positive in the interval $0 < x \le (1-\delta)/\e$.

\section{Conclusions}

We introduced and discussed in the present paper  kinetic models able to describe the statistical distribution of alcohol consumption in a multi-agent society. We succeeded to explain the emerging of a generalized Gamma distribution by resorting to the value function theory pioneered by Kahneman and Twersky \cite{KT,KT1}. In all cases, this macroscopic behavior is a consequence of the choice made at the microscopic level, choice that takes into account the essential features of the human behavior related to the phenomenon of alcohol consumption.  The kinetic modeling is similar to the one introduced in \cite{GT17}, subsequently generalized in \cite{GT18}, in which the human behavior was responsible of the formation of a macroscopic equilibrium in the form of a Lognormal distribution. Hence, the present results can be considered as an extension of the kinetic description of \cite{GT18}, which allows to classify, at a microscopic level, the main differences in the elementary interaction which produce a whole class of  generalized Gamma distributions, ranging from the classical Gamma density to the Lognormal one.  Well-known arguments of kinetic theory allow to model these phenomena by means of a Fokker--Planck equation with variable coefficients of diffusion and drift. Numerical experiments show the perfect agreement between the steady states of the kinetic model of Boltzmann type, and its continuous counterpart, represented by a Fokker--Planck type equation. 

We hope that the present analysis about the formation of generalized Gamma distribution in alcohol consumption can be of help to previous studies, like the ones presented in \cite{Keh,Par,Reh}, whose analysis was limited to Lognormal, Gamma and Weibull distributions. As remarked in \cite{Keh}, the accurate modeling of alcohol consumption as an upshifted distribution will provide public health decision-makers with accurate data to assess the
impact of alcohol consumption within and across countries and will aid in determining public health priorities and where to allocate resources. 

%%%%%%%%%%%%%%%%%%%%%%%%%%%%%%%%%%%%%%%%%%%%%%%%%%%%%%%%%%%%%%%%%%%%%%%%%%%%%%%

\section*{Acknowledgement} This work has been written within the
activities of GNFM group  of INdAM (National Institute of
High Mathematics), and partially supported by  MIUR project ``Optimal mass
transportation, geometrical and functional inequalities with applications''.
The research was partially supported by
the Italian Ministry of Education, University and Research (MIUR): Dipartimenti
di Eccellenza Program (2018--2022) - Dept. of Mathematics ``F.
Casorati'', University of Pavia.

%%%%%%%%%%%%%%%%%%%%%%%%%%%%%%%%%%%%%%%%%%%%%%%%%%%%%%%%%%%%%%%%%%%%%%%%%%%%%%%
\vskip 3cm
%\begin{table}
%\begin{tabular}{lll}
%\hline\noalign{\smallskip}
%\end{tabular}
%\end{table}

%\begin{acknowledgements}
%\end{acknowledgements}

\newpage

\end{document}